\documentclass[aps,prd,10pt,notitlepage,nofootinbib]{revtex4-1}

\usepackage[utf8]{inputenc}
\usepackage{amsmath,amssymb,amsfonts}
\usepackage{newtxtext,newtxmath}

\usepackage{bbold}
\usepackage{bm}
\usepackage{graphicx}
\usepackage[usenames,dvipsnames]{xcolor}
\usepackage{xcolor}
\usepackage[colorlinks=true,linkcolor=blue,urlcolor=blue,citecolor=blue]{hyperref}
\usepackage{slashed}
\usepackage[english]{babel}
\usepackage{dcolumn}
\usepackage{pifont}
\usepackage{dsfont,mathrsfs}
\usepackage{cancel}
\usepackage{bigints}
\usepackage{accents}
\usepackage{soul}
\usepackage{multirow}

\usepackage{amsmath, amssymb}
\usepackage{bbold}
\usepackage{makecell}
\usepackage{simpler-wick}
\usepackage{setspace}
\usepackage{orcidlink} %Added by Snigdha

\newcommand{\nn}{\nonumber}
\newcommand{\ket}[1]{\left|#1\right\rangle}

\newcommand{\ensembleaverage}[1]{\left\langle#1\right\rangle}
\newcommand{\MB}[1]{\left|#1\right|}
\newcommand{\FB}[1]{\left(#1\right)}
\newcommand{\SB}[1]{\left\{#1\right\}}
\newcommand{\TB}[1]{\left[#1\right]}

\newcommand{\scrL}{\mathscr{L}}
\newcommand{\scrM}{\mathscr{M}}

\newcommand{\munu}{{\mu\nu}}
\newcommand{\alphabeta}{{\alpha\beta}}

\newcommand{\IM}{\text{Im}}
\newcommand{\RE}{\text{Re}}

\newcommand{\wpv}{\omega_{\bm{p}}}
\newcommand{\wkv}{\omega_{\bm{k}}}

\newcommand{\domg}[1]{ \dfrac{d^3{#1}}{(2\pi)^3 2\omega_{\bm{#1}} }}

\newcommand{\tV}{\text{V}}
\newcommand{\tM}{\text{M}}
\newcommand{\tT}{\text{T}}
\newcommand{\tTM}{\text{TM}}

\definecolor{cerisepink}{rgb}{0.93, 0.23, 0.51}

\begin{document}
\title{Electrical conductivity and shear viscosity of a pion gas in a thermo-magnetic medium}

\author{Pallavi Kalikotay\orcidlink{0000-0003-0395-1445}$^{a,e}$}
\email{orionpallavi@gmail.com}

\author{Snigdha Ghosh\orcidlink{0000-0002-2496-2007}$^{b}$}
\email{snigdha.physics@gmail.com}
\email{snigdha.ghosh@bangla.gov.in}
\thanks{Corresponding Author}

\author{Nilanjan Chaudhuri\orcidlink{0000-0002-7776-3503}$^{c,f}$}
\email{sovon.nilanjan@gmail.com}
\email{n.chaudhri@vecc.gov.in}

\author{Pradip Roy$^{d,f}$}
\email{pradipk.roy@saha.ac.in}

\author{Sourav Sarkar\orcidlink{0000-0002-2952-3767}$^{c,f}$}
\email{sourav@vecc.gov.in}

\affiliation{$^a$Department of Physics, Kazi Nazrul University, Asansol - 713340, West Bengal, India}
\affiliation{$^b$Government General Degree College Kharagpur-II, Paschim Medinipur - 721149, West Bengal, India}
\affiliation{$^c$Variable Energy Cyclotron Centre, 1/AF Bidhannagar, Kolkata - 700064, India}
\affiliation{$^d$Saha Institute of Nuclear Physics, 1/AF Bidhannagar, Kolkata - 700064, India}
\affiliation{$^e$Department of Physics, Jadavpur University, Kolkata - 700032, West Bengal, India}
\affiliation{$^f$Homi Bhabha National Institute, Training School Complex, Anushaktinagar, Mumbai - 400085, India} 

%+++++++++++++++++++++++++++++++++++++++++++++++++++++++++++++++++++++++++++++++++++++++++++++++++++++++++++++++++++++

\begin{abstract}
We evaluate the electrical conductivity and shear viscosity of a interacting pion gas in a thermo-magnetic medium using the kinetic theory. The collision term of the relativistic Boltzmann transport equation in presence of background magnetic field is solved using the relaxation time approximation. The medium modified relaxation time is obtained from the corresponding in-medium $\pi\pi\rightarrow \pi\pi$ scattering cross-section calculated using the thermo-magnetic $\rho$ propagator. It is observed that the average relaxation time shows a $1/T^4$ variation with temperature for a fixed value of magnetic field. The relaxation time shows a mild oscillatory variation with respect to the magnetic field. It is also observed that the medium dependent scattering cross-section causes a considerable amount of influence on the electrical conductivity and shear viscosity compared to its vacuum counterpart.
\end{abstract}

\maketitle

\section{Introduction}
A strong magnetic field of the order of several $m_\pi^2$ or larger is produced in non-central relativistic heavy ion collisions (HIC) at RHIC and LHC~\cite{Kharzeev:2007jp,Skokov:2009qp}. The magnetic field $(eB)$ is comparable to the typical QCD scale and thus can have a direct influence on strongly interacting matter. The lifetime of the magnetic field is only a few fm/c. However, the finite electrical conductivity of the medium sustains the magnetic field for a longer time~\cite{Tuchin:2013apa,Gursoy:2014aka,Tuchin:2015oka,Das:2017qfi}. This magnetic field affects the evolution of the strongly interacting matter which has a noticeable effect on the dynamics and is reflected in the observables like charged hadron spectra and flow harmonics. Thus, a deeper understanding of the various aspects of strongly interacting matter in magnetic field is important to achieve the consistent dynamical modelling of the matter created in HICs.

The dynamical evolution of such a relativistic matter in presence of background magnetic field is studied using magnetohydrodynamics (MHD). Transport coefficients which are essential for describing the time evolution of strongly interacting matter are the key dynamical inputs to MHD. The estimation of transport coefficients in background magnetic field is not only important in the relativistic HICs but also in the context of other magnetized compact objects such as neutron stars and magnetars~\cite{Harutyunyan:2016rxm}. 

The transport coefficients are the measure of dissipative processes occurring in the strongly interacting matter. The dissipative processes are brought in by the collisions among the constituents. Thus the scattering cross-section is an important dynamical input to the transport coefficients. We know that in HICs, the hadronic phase attains a temperature in the range $100~\text{MeV} \lesssim T \lesssim 150 ~\text{MeV}$. Hence it becomes important to consider the thermal effects in the cross section calculations. Also it is observed in Ref.~\cite{Kalikotay:2020snc} that for a static medium the electrical conductivity of pion gas is estimated to be roughly $\sim$ 1-3 MeV which sustains the magnetic field of the order of $10^{-5}$GeV$^2$ for a time scale of 10 fm/c. Consequently, a weak magnetic field can be present in the hadronic phase of HICs, hence incorporation of magnetic field effects along with thermal effects can provide us a more realistic picture of HICs. Thus, we have considered the background magnetic field and thermal medium to calculate the transport coefficients in which thermo-magnetic effects are incorporated in the scattering cross-section. 

Substantial amount of works which involve the estimation of transport coefficients of quark gluon plasma (QGP) and hadrons in a background magnetic field such as electrical conductivity~\cite{Buividovich:2010tn,Pu:2014fva,Satow:2014lia,Gorbar:2016qfh,Harutyunyan:2016rxm,Kerbikov:2014ofa,Alford:2014doa,Nam:2012sg,Bandyopadhyay:2023lvk}, shear viscosity~\cite{Alford:2014doa,Huang:2011dc,Tuchin:2013ie,Finazzo:2016mhm,Ghosh:2022xtv,Ghosh:2020wqx}, bulk viscosity and thermal conductivity~\cite{Dash:2020vxk,Das:2019pqd} are available in the literature. In Ref.~\cite{Das:2019pqd}, the transport coefficients of a hadron resonance gas has been evaluated using a constant cross-section whereas in Ref.~\cite{Dash:2020vxk}, the relaxation time has been taken as a parameter to evaluate the necessary transport coefficients. 

The evaluation of electrical conductivity as mentioned earlier will give us an insight on the lifetime of the magnetic field generated in HIC whereas the shear viscosity provides a useful signature of phase transition between deconfined quark matter and confined hadronic matter~\cite{Niida:2021wut}. Also, shear viscosity calculation becomes important because in the recent past a finite yet small shear viscosity over entropy density ratio namely $\eta/s$ could explain the elliptic flow data in viscous hydrodynamic simulations of charged hadrons in Au+Au collisions at $\sqrt{s}=200$ GeV per nucleon~\cite{Luzum:2008cw}. Since pions are the most abundant species produced in the hadronic phase, in this work we will evaluate the electrical conductivity and shear viscosity of a pion gas in a thermo-magnetic medium. Incorporation of the effect of magnetic field along with thermal effects on the relaxation time is the novelty of this work. 

This article is organized as follows. Sec.~\ref{sec.for} along with its two subsections deals with the evaluation of electrical conductivity and shear viscosity using the Boltzmann transport equation (BTE) in presence of background magnetic field within the ambit of relaxation time approximation. In Sec.~\ref{sec.tau}, we discuss the the evaluation of the relaxation time of pions in a thermo-magnetic medium which is followed with a discussion on numerical results in Sec.~\ref{sec.result}. Finally, we conclude with the summary of our work in Sec.~\ref{sec.sum}. Some of the calculational details are provided in the Appendices.
%~~~~~~~~~~~~~~~~~~~~~~~~~~~~~~~~~~~~~~~~~~~~~~~~~~~~~~~~~~~~~~~~~~~~~~~~~~~~~~~~~~~~~~~~~~
%
\section{Formalism} \label{sec.for}
In order to derive the expression for electrical conductivity and shear viscosity in a magnetized medium, we start with the standard expression of relativistic BTE in background electromagnetic field (characterized by the field strength tensor $F^\munu$), which is given by~\cite{DeGroot:1980dk}
\begin{eqnarray}
	p^\mu \partial_\mu f
%	(\wpv)
	 + q F^{\mu\nu} p_\nu \frac{\partial f}{\partial p^\mu} = C[f] \label{BTE_magnetic}
\end{eqnarray}
where, $f$ is the single particle phase space distribution function, $q$ is the electric charge of the particle and the metric tensor with signature $g^\munu=\text{diag}(1,-1,-1,-1)$ has been used throughout the work. The collision term on the right hand side of Eq.~\eqref{BTE_magnetic} will be solved using the relaxation time approximation whereby we consider $C[f]=- \frac{\delta f}{\tau} $, in which $\delta f$ is the deviation function and $\tau$ is the relaxation time. The deviation function $\delta f$ encapsulates the dissipative processes occurring in the system brought in by the collisions among the constituents. In next two subsections, we will obtain the expressions of conductivities and shear viscosities by chosing appropriate form of the deviation function $\delta f$.
%~~~~~~~~~~~~~~~~~~~~~~~~~~~~~~~~~~
\subsection{Electrical conductivity in magnetic field}
For solving Eq.~\eqref{BTE_magnetic} for the electrical and Hall conductivity, we consider the following functional form for the deviation function
\begin{eqnarray}
	\delta f = -\phi \frac{\partial f_0}{\partial \wpv} = -(\bm{p}\cdot\bm{\Xi}(\wpv))\frac{\partial f_0}{\partial \wpv}, \label{tmp.1}
\end{eqnarray}
where the equilibrium distribution function is given by 
$f_0(\wpv) = \frac{1}{e^{\wpv/T} - 1}$ in which $T$ is the temperature and $\wpv=\sqrt{\bm{p}^2 + m_\pi^2}$ is the single particle pion energy. In Eq.~\eqref{tmp.1}, the vector quantity $\bm\Xi(\wpv)$ accounts for the dissipative effects brought in by the electric and magnetic field and is given as
\begin{eqnarray}
	\bm{\Xi}(\wpv) = \alpha \hat{\bm{e}} + \beta \hat{\bm{b}} + \gamma (\hat{\bm{e}}\times \hat{\bm{b}}),
\end{eqnarray}
where $\hat{\bm{e}}$ and $\hat{\bm{b}}$ are the unit vectors along the directions of electric field $\bm{E}=|\bm{E}|\hat{\bm{e}}$ and magnetic field $\bm{B}=|\bm{B}|\hat{\bm{b}}$ respectively.
The unknown coefficients $\alpha,~\beta$ and $\gamma$ can be solved by substituting the expression of $\delta f$ from Eq.~\eqref{tmp.1} into the BTE. Having obtained the deviation function $\delta f$ (or equivalently $\phi$), the conductivity tensor $\sigma^{ij}$ can be extracted from the definition of the macroscopic current density $j^i$, which is given by
\begin{eqnarray}
	j^i = \sigma^{ij} E^j = 2\int\!\frac{d^3 p}{(2\pi)^3}e v^i \phi \frac{\partial f_0}{\partial \wpv} \label{j_i}.
\end{eqnarray} 
The tensorial decomposition of the conductivity tensor $\sigma^{ij}$ can be taken as~\cite{Harutyunyan:2016rxm}
\begin{eqnarray}
	\sigma^{ij} = \delta^{ij} \sigma_0 -\epsilon^{ijk} b^k \sigma_1 + b^i b^j \sigma_2 \label{sigma.ij}
\end{eqnarray}
where, $\sigma_0$ is the electrical conductivity in presence of the magnetic field, $\sigma_1$ is the Hall conductivity and $(\sigma_0 + \sigma_2)$ is the electrical conductivity in absence of magnetic field. The general expression for different components $\sigma_n$ (with $n=0,1,2$) of electrical conductivity in presence of magnetic field is found by substituting the expression of $\phi$ into Eq.~\eqref{j_i} and making use of Eq.~\eqref{sigma.ij} as done in Ref.~\cite{Kalikotay:2020snc}, to obtain
\begin{eqnarray}
	\sigma_0 &=& \frac{g_\pi e^2}{3T} \int\!\! \frac{d^3 p}{(2\pi)^3} \frac{\bm{p}^2}{\wpv^2} \frac{\tau}{1+ (\omega_c \tau)^2} f_0(\wpv) (1+f_0(\wpv)) \label{sigma0},\\
	\sigma_1 &=& \frac{g_\pi e^2}{3T} \int\!\! \frac{d^3 p}{(2\pi)^3} \frac{\bm{p}^2}{\wpv^2} \frac{\tau (\omega_c \tau)}{1+ (\omega_c \tau)^2} f_0(\wpv) (1+f_0(\wpv)), \label{sigma1}\\
	\sigma_2 &=& \frac{g_\pi e^2}{3T} \int\!\! \frac{d^3 p}{(2\pi)^3} \frac{\bm{p}^2}{\wpv^2} \frac{\tau (\omega_c \tau)^2}{1+ (\omega_c \tau)^2} f_0(\wpv) (1+f_0(\wpv)) \label{sigma2}
\end{eqnarray}
where $g_\pi$ is the charged pion degeneracy, $\omega_c = \frac{|q\bm{B}|}{\wpv}$ is the cyclotron frequency. The details of the calculation can be found in Ref.~\cite{Kalikotay:2020snc}.
%~~~~~~~~~~~~~~~~~~~~~~~~~~~~~~~~~~~~~~~~~~~~~~~~~~~~~~~~~~~~
\subsection{Shear viscosity in magnetic field}
The velocity gradients in the strongly interacting matter give rise to the shear viscosity in the medium. In order to extract the shear viscosity coefficients in presence of external magnetic field, Eq.~\eqref{BTE_magnetic} can be reduced to (taking $\bm{E}=\bm{0}$)
\begin{eqnarray}
	- \frac{1}{T} p_i p_j V_{ij} f_0(1+f_0) + q (\bm{v}\times \bm{B})\cdot \frac{\partial (\delta f)}{\partial \bm{p}} = - \frac{\delta f}{\tau} \label{BTE_shear}
\end{eqnarray}
where $\bm{v} = \bm{p}/\wpv$, and the tensor $V_{ij} = \frac{1}{2}\left( \partial_i v_j + \partial_j v_i \right)$ contains the velocity gradients. 
Eq.~(\ref{BTE_shear}) can be further reduced to
\begin{eqnarray}
	- \frac{1}{T} p_i p_j V_{ij} f_0(1+f_0) + \frac{q|\bm{B}|}{\wpv}\epsilon_{ijk} v_j b_k \frac{\partial (\delta f)}{\partial {v_i}} = - \frac{\delta f}{\tau}. \label{BTE_shear1}
\end{eqnarray}
In order to solve Eq.~\eqref{BTE_shear1} for shear viscosities, the deviation function $\delta f$ containing the velocity gradients is taken to be of the form 
\begin{eqnarray}
	\delta f = \sum_{n=0}^{n=4} g_n V_{ij}^n v_i v_j \label{delta_shear}
\end{eqnarray}
where $g_n$'s are the coefficients to be determined, and the velocity gradients $V_{ij}^n$'s in presence of magnetic field are constructed using the available vector $b_i$, the unit tensor $\delta_{ij}$ and the Levi-Civita symbol $\epsilon_{ijk}$ as follows~\cite{Landau}
\begin{eqnarray}
	V_{ij}^0 &=& (3b_i b_j - \delta_{ij}) \left(b_k b_l V_{kl} - \frac{1}{3} \bm{\nabla}\cdot \bm{V}\right),  \label{V0}\\
	V_{ij}^1 &=& 2 V_{ij} + \delta_{ij} V_{kl} b_k b_l - 2 V_{ik} b_k b_j - 2 V_{jk} b_k b_i + (b_{ij} - \delta_{ij}) \bm{\nabla}\cdot \bm{V} + b_i b_j V_{kl} b_k b_l,  \label{V1} \\
	V_{ij}^2 &=& 2 \left( V_{ik} b_j b_k + V_{jk} b_i b_k - 2 b_i b_j V_{kl}b_k b_l  \right), \label{V2}\\
	V_{ij}^3 &=& V_{ik}b_{jk} + V_{jk} b_{ik} - V_{kl} b_{ik} b_j b_l - V_{kl} b_{jk} b_i b_l, \label{V3}\\
	V_{ij}^4 &=& 2 \left( V_{kl} b_{ik} b_j b_l + V_{kl} b _{jk} b_i b_l \right), \label{V4}
\end{eqnarray}
in which $b_{ij}=\epsilon_{ijk} b_k$. In order to calculate the transverse shear viscosity, we use $\bm{\nabla}\cdot \bm{V} = 0$ and $V_{kl} b_k b_l=0$, as a result of which $V_{ij}^0=0$. We now proceed to solve Eq.~\eqref{BTE_shear1} for the deviation function $\delta f$ for which we substitute Eq.~\eqref{delta_shear} into Eq.~\eqref{BTE_shear1} and make use of Eqs.~\eqref{V0}-\eqref{V4}, to obtain (after some algebra): 
\begin{eqnarray}
	\frac{\wpv}{T} v_i v_j V_{ij} f_0(1+f_0) &=& 
	2\omega_c g_1 \left[ 2V_{ik} b_{ij} v_j v_k - 2 V_{ik} b_{ij} b_k v_j (\bm{b}\cdot\bm{v}) \right]  + 2\omega_c g_2 \left[ 2V_{ik} b_{ij}b_k v_j  (\bm{b}\cdot\bm{v})  \right] \nonumber \\
	&& +~ 2\omega_c g_3 \left[ 2 V_{ij} v_i v_j - 4  V_{ij}b_j v_i (\bm{b}\cdot\bm{v})   \right]  +  2\omega_c g_4 \left[2 V_{ij}b_j v_i (\bm{b}\cdot\bm{v}) \right] \nonumber \\
	&& +~  \frac{g_1}{\tau} \left[ 2 V_{ij} v_i v_j - 4  V_{ij}b_j v_i (\bm{b}\cdot\bm{v})  \right] +  \frac{g_2}{\tau} \left[ 4  V_{ij}b_j v_i (\bm{b}\cdot\bm{v})  \right] \nonumber \\
	&& +~ \frac{g_3}{\tau} \left[ 2V_{ik} b_{jk} v_i v_j - 2V_{kj} b_{ik} b_j v_i (\bm{b}\cdot\bm{v})  \right]  + \frac{g_4}{\tau} \left[ 4V_{kj} b_{ik} b_j v_i (\bm{b}\cdot\bm{v})\right]. \label{BTE_shear2}	
\end{eqnarray}
Comparing the coefficients of tensor structures in Eq.~\eqref{BTE_shear2} on both sides, we arrive at the following set of linear equations in $g$'s
\begin{eqnarray}
	2\omega_c g_3 + \frac{g_1}{\tau} &=& \frac{\wpv}{2T} f_0 (1+f_0), \label{shear_compare1}\\
	g_3 - 2 \omega_c \tau g_1 &=& 0, \label{shear_compare2}\\
	2 \omega_c g_1 - 2\omega_c g_2 - \frac{g_3}{\tau} + \frac{2g_4}{\tau} &=& 0,  \label{shear_compare3}\\
	2 \omega_c g_3 - \omega_c g_4 + \frac{g_1}{\tau} - \frac{g_2}{\tau} &=& 0.  \label{shear_compare4}
\end{eqnarray} 
By solving Eqs.~\eqref{shear_compare1}-\eqref{shear_compare4}, the unknown quantities $g_n$ with $n=1,2,3,4$ is obtained as
\begin{eqnarray}
	g_1 &=& \frac{\wpv}{2T} ~\frac{\tau}{\left[ 1 + 4 (\omega_c\tau)^2 \right]} f_0 (1+f_0), \label{g1}\\
	g_2 &=& \frac{\wpv}{2T}~ \frac{\tau}{\left[ 1 + (\omega_c\tau)^2 \right]} f_0 (1+f_0), \label{g2}\\
	g_3 &=& \frac{\wpv}{2T} ~\frac{2\omega_c \tau^2}{\left[ 1 + 4 (\omega_c\tau)^2 \right]} f_0 (1+f_0),  \label{g3}\\
	g_4 &=& \frac{\wpv}{2T} ~\frac{\omega_c \tau^2}{\left[ 1 + (\omega_c\tau)^2 \right]} f_0 (1+f_0). \label{g4}
\end{eqnarray}

Viscosity is characterized by the non-uniformity in the fluid flow and this information is carried by the deviation function. Therefore, having calculated the deviation function $\delta f$ (or equivalently the $g$'s), the transverse shear viscous coefficients can be extracted from the definition of the macroscopic momentum flux density tensor or the momentum flow tensor $\pi^{ij}$, which reads
\begin{eqnarray}
	\pi_{ij} = \int\!\! \frac{d^3p}{(2\pi)^3} v_i v_j \wpv \delta f = \sum_{n=0}^{n=4} \int\!\! \frac{d^3p}{(2\pi)^3} \wpv g_n v_i v_j v_k v_l V_{kl}^n 
	 \label{pi_ij_delta}
\end{eqnarray}
where in the last step, $\delta f$ has been substituted from Eq.~\eqref{delta_shear}. Alternatively, the general tensorial decomposition of $\pi^{ij}$ in terms of the velocity gradients can be written as
\begin{eqnarray}
	\pi_{ij} = \sum_{n=0}^{n=4} \eta_n V_{ij}^n  \label{pi_ij_eta}
\end{eqnarray}
where $\eta_n$'s are the shear viscosity coefficients in presence of external magnetic field. Making use of $V_{ij}^0=0$ for transverse shear viscous coefficients and comparing Eq.~\eqref{pi_ij_delta} and Eq.~\eqref{pi_ij_eta}, we get after some algebra
\begin{eqnarray}
	\eta_n = \frac{2}{15} \int\!\! \frac{d^3p}{(2\pi)^3} \wpv g_n v^4 ~~~,~~~~~~~n =1,2,3,4. \label{eta_n}
\end{eqnarray}
Finally, substituting Eqs.~\eqref{g1}-\eqref{g4} into Eq.~\eqref{eta_n}, we obtain the following final expressions for the shear viscosity coefficients
\begin{eqnarray}
	\eta_1 &=& \frac{g_\pi}{15T} \int\!\! \frac{d^3 p}{(2\pi)^3} \frac{\bm{p}^4}{\wpv ^2}  \frac{\tau}{1  + (2\tau \omega_c)^2} f_0(\wpv) \{1+f_0(\wpv)\}, \label{eta_1} \\
	\eta_2 &=& \frac{g_\pi}{15T} \int\!\! \frac{d^3 p}{(2\pi)^3} \frac{\bm{p}^4}{\wpv ^2}  \frac{\tau}{1  + (\tau \omega_c)^2} f_0(\wpv) \{1+f_0(\wpv)\}, \label{eta_2}  \\
	\eta_3 &=& \frac{g_\pi}{15T} \int\!\! \frac{d^3 p}{(2\pi)^3} \frac{\bm{p}^4}{\wpv ^2}  \frac{\tau^2 \omega_c}{\frac{1}{2} + 2(\tau \omega_c)^2} f_0(\wpv) \{1+f_0(\wpv)\}, \label{eta_3} \\
	\eta_4 &=& \frac{g_\pi}{15T} \int\!\! \frac{d^3 p}{(2\pi)^3} \frac{\bm{p}^4}{\wpv ^2}  \frac{\tau^2 \omega_c}{1  + (\tau \omega_c)^2} f_0(\wpv) \{1+f_0(\wpv)\}, \label{eta_4} 
\end{eqnarray}
where, $\eta_{1}$ and $\eta_2$ are the shear viscosity coefficients in presence of magnetic field and $\eta_3$ and $\eta_4$ are Hall type shear viscosities as they vanish for vanishing magnetic field whereas $\eta_1 = \eta_2 =\eta $ for vanishing magnetic field. 
%~~~~~~~~~~~~~~~~~~~~~~~~~~~~~~~~~~~~~~~~~~~~~~~~~~~~~~~~~~~~~~~~~~~~~~~~~~~~~~~
\section{Relaxation time in thermo-magnetic medium} \label{sec.tau}
Let us now calculate the relaxation time $\tau$ appearing in the electrical conductivity and shear viscosity expression. 
For the $2\to2$ scattering process $\pi (p) + \pi (k) \rightarrow \pi(p') + \pi(k')$, the inverse of the relaxation time $\tau$ is given by
\begin{eqnarray}
	\frac{1}{\tau(p)} = \dfrac{g}{4\omega_{p}} \int\!\!\!\int\!\!\!\int\!\!\! \domg{k} \domg{p'} \domg{k'}(2\pi)^4 \delta ^4 \FB{p + k - p' -k' }\MB{ \scrM }^2  
	\frac{f_0^{\bm{p}} (1+f_0^{\bm{k'}}) (1+f_0^{\bm{p'}}) } { (1+f_0^{\bm{k}}) } 
	\label{tau_ab} 
\end{eqnarray}
where, $\scrM$ is the scattering amplitude, $f_0^{\bm{p}} = f_0(\wpv)$, $f_0^{\bm{k'}} = f_0(\omega_{\bm{k'}})$ etc., and $g=3$ is the degeneracy of the pions $(\pi^{\pm},\pi^0)$.
Integrating over the momenta $d^3k'$ and $d^3p'$, we get 
\begin{equation}
	\frac{1}{\tau(p)} = \frac{g}{2} \int\!\! \dfrac{d^3 k}{(2\pi)^3} \FB{\sigma ~ v_{\rm rel}} 
	f_0(\omega_k)\SB{1 + f_0(\omega_k)} \label{tau_k}
\end{equation} 
where, $\sigma$ is the total $\pi\pi \rightarrow \pi\pi$ scattering cross section, and $v_\text{rel}$ is the relative velocity between the initial state particles. We will now calculate the $\pi\pi \rightarrow \pi\pi$ cross section in a thermo-magnetic medium mediated by the vector meson $\rho$. The interaction Lagrangian (density) used is~\cite{Krehl:1999km}
\begin{eqnarray}
	\scrL_\text{int} = - g_{\rho\pi\pi}\partial_\mu \bm{\rho}_\nu \cdot (\partial^\mu \bm{\pi}\times\partial^\nu\bm{\pi})  \label{eq.lagrangian}  
\end{eqnarray}
where, $\bm{\rho}_\nu$ and $\bm{\pi}$ are the iso-vector fields corresponding to $\rho$ and $\pi$ mesons respectively, and the coupling constant $g_{\rho\pi\pi}= 20.72$ GeV$^{-2}$ is estimated from the experimental vacuum decay width of $\rho\to\pi\pi$ which is about $\Gamma_{\rho\to\pi\pi} \simeq 150$ MeV. It is useful to consider the isospin basis, so that the isospin averaged total cross section can be written as  
\begin{eqnarray}
	\sigma = \frac{1}{64 \pi^2 s} \int\!\! d\Omega~ \FB{\frac{5}{9} \MB{\scrM_2}^2 + \frac{1}{3} \MB{\scrM_1}^2 + \frac{1}{9} \MB{\scrM_0}^2} \label{xsection}
\end{eqnarray}
where, $\scrM_I$ is the invariant amplitude in isospin channel(s) corresponding to the composite pionic isospin state(s) $\ket{I,I_z}$ having total isospin $I$ (we note that, amplitudes are independent of the third component $I_z$ of the isospin vector $I$). Using Eq.~\eqref{eq.lagrangian}, the explicit expressions of $\scrM_I$'s come out to be~\cite{Mitra:2012jq}
\begin{eqnarray}
	\scrM_{0} &=& g_{\rho\pi\pi}^2\left[2\left(\frac{s-u}{t- m_\rho^2}\right)+2\left(\frac{s-t}{u-m_\rho^2}\right)\right], \\
	\scrM_{1} &=& g_{\rho\pi\pi}^2\left[ 2\left(\frac{t-u}{s-m_\rho^2 - \frac{1}{3} g_\munu\Pi^\munu(\sqrt{s},\bm{0};T,B)}\right)+\left(\frac{s-u}{t-m_\rho^2}\right)
	-\left(\frac{s-t}{u-m_\rho^2}\right)\right], \label{M1}\\
	\scrM_{2} &=& g_{\rho\pi\pi}^2\left[-\left(\frac{s-u}{t-m_\rho^2}\right)-
	\left(\frac{s-t}{u-m_\rho^2}\right) \right], \\
\end{eqnarray}
where, $s,t,u$ are the Mandelstam variables, $m_\rho$ is the mass of the $\rho$ meson, and $\Pi^\munu(q^0,\bm{q};T,B)$ is the thermo-magnetic self energy function of $\rho$-meson with momentum $q$.
\begin{figure}[h]
	\includegraphics[angle=0, scale=0.35]{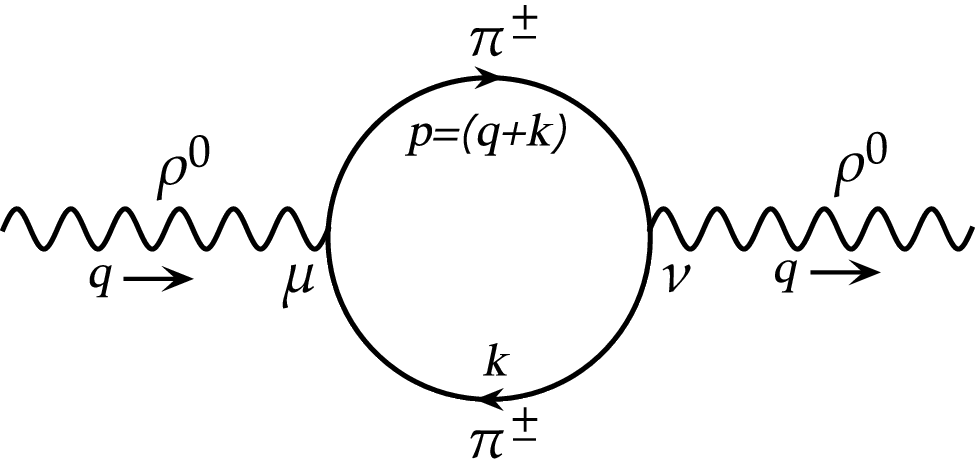}  ~~~~~~~
	\includegraphics[angle=0, scale=0.35]{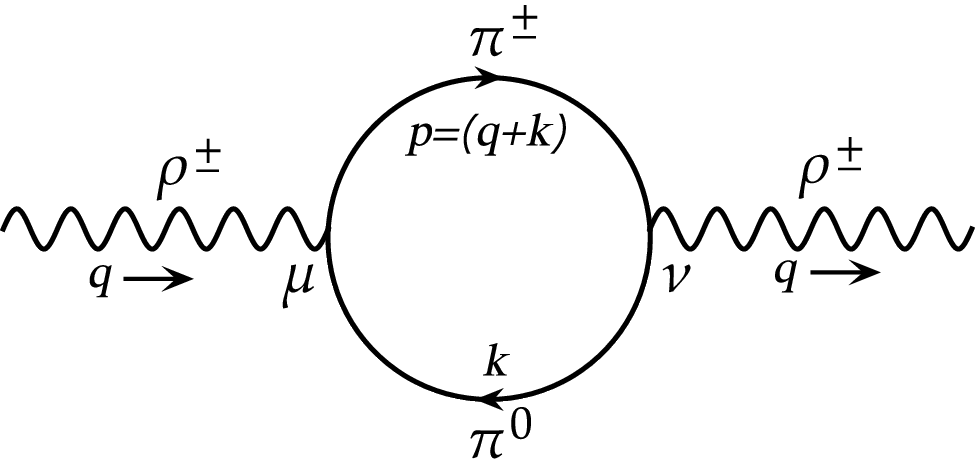} 
	\caption {Feynman diagrams for the one-loop self energy of neutral and charged $\rho$ meson.} 
	\label{self_energy}
\end{figure}

We will now evaluate the one-loop self-energies of the $\rho$ meson in a thermo-magnetic medium. Fig.~\ref{self_energy} shows the Feynman diagrams for the one-loop self-energies of neutral and charged rho-mesons originating from the Lagrangian in Eq.~\eqref{eq.lagrangian}. Unlike the zero magnetic field case, the self-energies of $\rho^0$ and $\rho^{\pm}$ become unequal in the presence of an external magnetic field. For calculating the thermo-magnetic self-energy of $\rho$-meson, we will use the real time formulation of thermal field theory in a magnetic background, in which the self-energy becomes $2\times2$ matrix~\cite{Bellac:2011kqa,Mallik:2016anp}. Applying the finite temperature Feynman rules to Fig.~\ref{self_energy}, the 11-components of the one-loop real time $\rho^0$ and $\rho^\pm$ meson self-energy matrices come out to be~\cite{Ghosh:2017rjo}
\begin{eqnarray}
	\Pi^{\mu\nu}_{11,0}(q;T,eB) &=& i \int\! \frac{d^4k}{(2\pi)^4}N^{\mu\nu}(q,k)D^\text{mag}_{11}(k)D^\text{mag}_{11}(p=q+k), \label{Pi.11.0} \\
	\Pi^{\mu\nu}_{11,\pm}(q;T,eB) &=& i \int\! \frac{d^4k}{(2\pi)^4}N^{\mu\nu}(q,k)D_{11}(k)D^\text{mag}_{11}(p=q+k) \label{Pi.11.pm}
\end{eqnarray}
where $N^{\mu\nu} (q,k) = g_{\rho\pi\pi}^2 \left[ q^4 k^{\mu} k^{\nu} + (q\cdot k)^2 q^{\mu} q^{\nu} - q^2 (q\cdot k) (q^{\mu} k^{\nu} + k^{\mu} q^{\nu}) \right]$, $D_{11}(k)$ and $D^\text{mag}_{11}(k)$ are the 11-component of the real time neutral and charged pion propagators respectively in a magnetic background (the $\pi^0$ propagator is not affected by the magnetic field). The explicit expressions of the $\pi^0$ and $\pi^\pm$ propagators read~\cite{Ghosh:2017rjo,Ayala:2016awt} 
	\begin{eqnarray}
	D_{11}(k)&=& \TB{\frac{-1}{k^2-m_\pi^2+i\epsilon}+2\pi i\eta(k^0)\delta(k^2-m_\pi^2)}, \label{} \\
	D^\text{mag}_{11}(k)&=&\sum_{n=0}^{\infty}2(-1)^ne^{-\alpha_k}L_n(2\alpha_k)\TB{\frac{-1}{k_\parallel^2-m^2_n+i\epsilon}+2\pi i\eta(k^0)\delta(k_\parallel^2-m^2_n)}\label{}
\end{eqnarray}
where, we have considered external magnetic field $\bm{B}=B\hat{\bm{z}}$ along the positive-$\hat{\bm{z}}$ direction, $\eta(x)=\Theta(x)f_0(x)+\Theta(-x)f_0(-x)$, $n$ is the Landau level index, $\alpha_k=-k_\perp^2/eB>0$, $e$ is the electric charge of a proton, $L_l(z)$ is Laguerre polynomial of order $l$, $m_n=\sqrt{m_\pi^2+(2n+1)eB}$,  $p_{\parallel,\perp}^\alpha = g_{\parallel,\perp}^\alphabeta p_\beta$ with $g_\parallel^\munu=\text{diag}(1,0,0,-1)$ and $g_\perp^\munu=\text{diag}(0,-1,-1,0)$.

Having obtained the 11-components of the self-energies, the analytic thermo-magnetic self-energy function $\Pi^\munu(q;T,B)$ appearing in the expression of scattering amplitude in Eq.~\eqref{M1} can be obtained via relations~\cite{Mallik:2016anp,Bellac:2011kqa} $\IM\Pi^{\mu\nu}(q)=\tanh\FB{\frac{|q^0|}{2T}}\IM\Pi^{\mu\nu}_{11}(q)$ and $\RE\Pi^{\mu\nu}(q)=\RE\Pi^{\mu\nu}_{11}(q)$. The imaginary parts of the self-energies can be simplified to obtain
\begin{eqnarray}
	\IM{\Pi}^{\mu\nu}_0(q^0,\bm{q}=\bm{0}) &=& \frac{-1}{4q^0}\sum\limits_{l=0}^{\infty} \sum\limits_{n=0}^{\infty}\frac{1}{{k}_z^\prime} 
	\left[ U_{1,nl}^\munu( q^0,k_z^\prime)\Theta (q^0-m_n-m_l) + U_{2,nl}^\munu( q^0,k_z^\prime)\Theta(-q^0-m_n-m_l) \right. \nonumber \\
	&& \hspace{2cm} \left. +~L_{1,nl}^\munu( q^0,k_z^\prime) \Theta\left\{-q^0-\min(m_n-m_l,0)\right\} \Theta\left\{\max(m_n-m_l,0)+q^0\right\} \nonumber  \right. \\
	&& \hspace{2cm} \left. +~ L_{2,nl}^\munu( q^0,k_z^\prime) \Theta\left\{q^0-\min(m_n-m_l,0)\right\} \Theta\left\{\max(m_n-m_l,0)-q^0\right\} \right], \label{impi0} \\
%\end{eqnarray}
%\begin{eqnarray}
%
\IM{\Pi}^{\mu\nu}_\pm (q^0,\bm{q}=\bm{0}) &=& \frac{-\text{sign}(q^0)}{16\pi}\sum\limits_{l=0}^{\infty} \sum_{s \in \{\pm\} }^{} \Bigg[
\int\limits_{\omega_0}^{\omega_-}\frac{d\omega_{\bm{k}}}{|\bm{k}|\cos\theta_0^+}U_{1,l}^\munu\big(q^0,|\bm{k}|,s\cos\theta_0^+\big) \Theta( q^0-m_\pi-m_l)  \nonumber \\
&&  + \int\limits_{-\omega_0}^{-\omega_+}\frac{d\omega_{\bm{k}}}{|\bm{k}|\cos\theta_0^-} U_{2,l}^\munu\big(q^0,|\bm{k}|,s\cos\theta_0^-\big) \Theta(-q^0-m_\pi-m_l)  \nonumber \\ 
&&  + \int\limits_{-\omega_-}^{-\omega_0}\frac{d\omega_{\bm{k}}}{|\bm{k}|\cos\theta_0^-} L_{1,l}^\munu\big(q^0,|\bm{k}|,s\cos\theta_0^-\big) \Theta(-q^0-m_\pi+m_l)\Theta(q^0)  \nonumber \\
&&   + \int\limits_{\omega_+}^{\omega_0}\frac{d\omega_{\bm{k}}}{|\bm{k}|\cos\theta_0^+} L_{2,l}^\munu\big(q^0,|\bm{k}|,s\cos\theta_0^+\big) \Theta(q^0-m_\pi+m_l)\Theta(-q^0)\Bigg]\label{impipm} 
\end{eqnarray}
where ${k}_z^\prime = \frac{1}{2q^0}\lambda^{1/2}(q_0^2, m_l^2, m_n^2)$, $\lambda(x,y,z)=x^2+y^2+z^2-2xy-2yz-2zx$ is the K\"all\'en function, $\omega_0 = \frac{1}{2q^0}(q_0^2 + m_\pi^2 - m_l^2)$, $\omega_{\pm}=q^0 \pm m_l$, and $\cos\theta_0^\pm = \frac{1}{|\bm{k}|}\sqrt{(q^0 \mp \omega_k)^2 -m_l^2}$. 
The explicit expressions of the tensors $U_{1,nl}^\munu$, $U_{2,nl}^\munu$, $L_{1,nl}^\munu$, and $L_{2,nl}^\munu$ appearing in Eq.~\eqref{impi0} for the neutral rho-meson and $U_{1,l}^\munu$, $U_{2,l}^\munu$, $L_{1,l}^\munu$, and $L_{2,l}^\munu$ appearing in Eq.~\eqref{impipm} for the charged rho-meson are provided in Appendix~\ref{app.tensors}.
The imaginary parts of the $\rho^0$ as well as $\rho^\pm$ self energies have contributions from four terms (symbolically the $U_1$, $U_2$, $L_1$ and $L_2$) containing a number of step functions. These step functions represent the branch cuts of the self-energy function in the complex $q^0$ plane. They are termed as Unitary-I, Unitary-II, Landau-I and Landau-II cuts respectively as they appear in Eqs.~\eqref{impi0} and \eqref{impipm}. These cuts physically correspond to the kinematically allowed scattering and decay processes in the thermomagnetic medium. The details of the calculation and an analysis of the analytic structure of the self-energy function can be found in Ref.~\cite{Ghosh:2017rjo}. Having obtained the thermo-magnetic self-energy functions $\Pi^{\mu\nu}_{0,\pm}$ of $\rho^0$ and $\rho^\pm$, we substitute them into the expression of scattering amplitude in Eq.~\eqref{M1} to obtain the isospin averaged $\pi\pi\to\pi\pi$ cross-section from Eq.~\eqref{xsection} for a $\rho^0$ or $\rho^\pm$ exchange. 
%
%\fixme{
%We also point out that, while calculating the cross-section, we have approximated that the $\rho^\pm$ propagation is not directly/primarily affected by the external magnetic field; and the magnetic modification to the $\rho^\pm$ propagation occurs only due to its fluctuation into pion pair which are influenced by the external magnetic field. In other words, the Landau quantization of the charged rho has not been considered in this work. This approximation is well justified due to the fact that $m_\rho^2 \gg eB$ for the typical magnetic field strength in the hadronic phase of HICs, and hence the Landau levels of $\rho^\pm$ are too closely spaced and can almost be treated as a continuum.  
%}
%
%~~~~~~~~~~~~~~~~~~~~~~~~~~~~~~~~~~~~~~~~~~~~~~~~~~~~~~~~~~~~~~
\section{Numerical Results \& Discussions} \label{sec.result}
%
%We begin this section by evaluating the numerical results for a generic isospin averaged $\pi\pi$ cross-section mediated by neutral and charged $\rho$ mesons. The numerical results of the cross-section has been computed in various media such as vacuum (V), thermal medium (T), magnetic medium (M), and thermo-magnetic medium (TM). The neutral and charged $\rho$ mesons are affected differently in the presence of magnetic field as evident from the analytical calculation in the previous section. 

We begin this section by evaluating the numerical results for a generic isospin averaged $\pi\pi$ cross-section mediated by neutral and charged $\rho$ mesons.  The neutral and charged $\rho$ mesons are affected differently in the presence of magnetic field as evident from the analytical calculation in the previous section. 
\\

% These media are denoted respectively as V, T, M and TM. 
\begin{figure}[h]
	\includegraphics[angle=-90, scale=0.3]{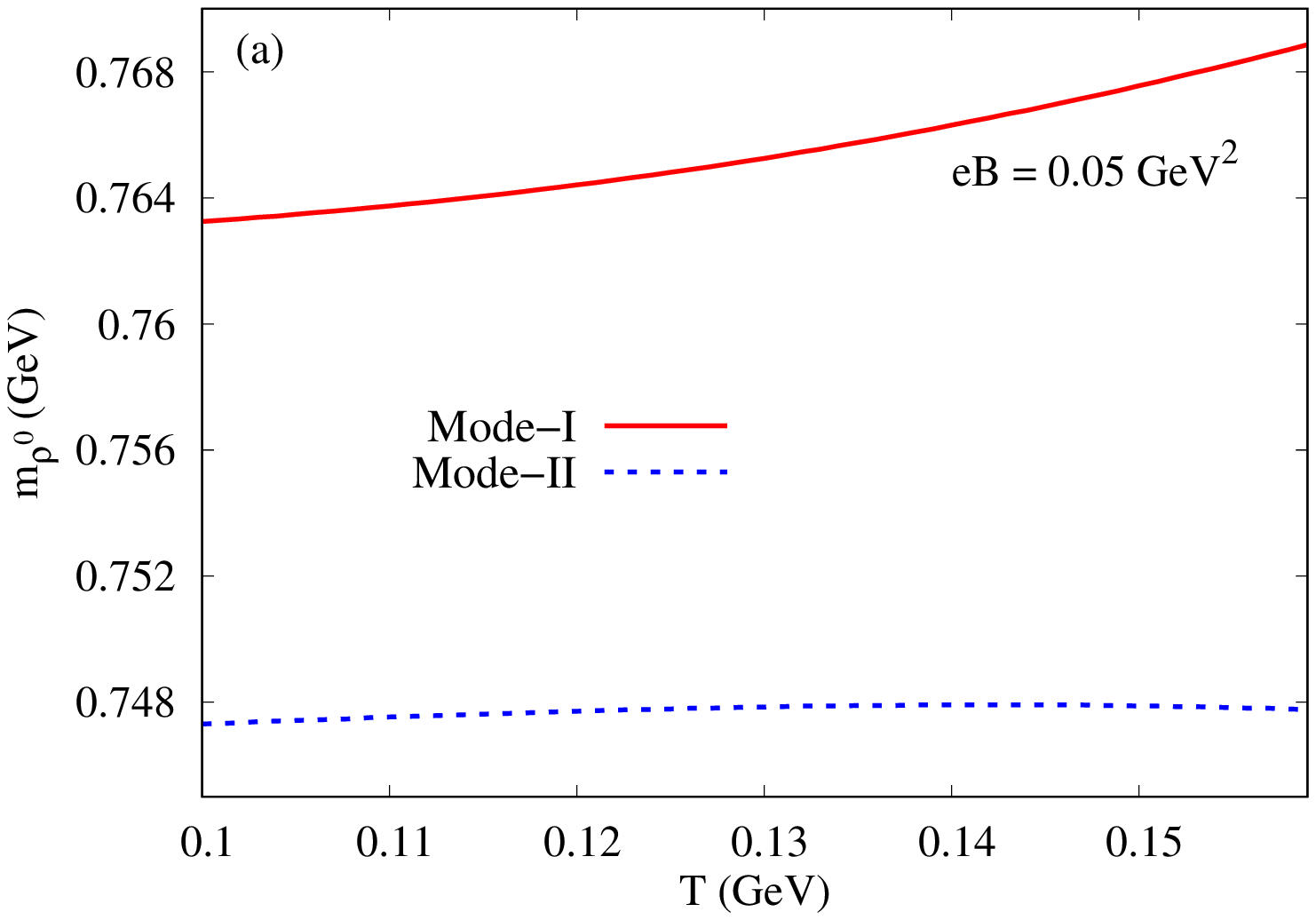}  
	\includegraphics[angle=-90, scale=0.3]{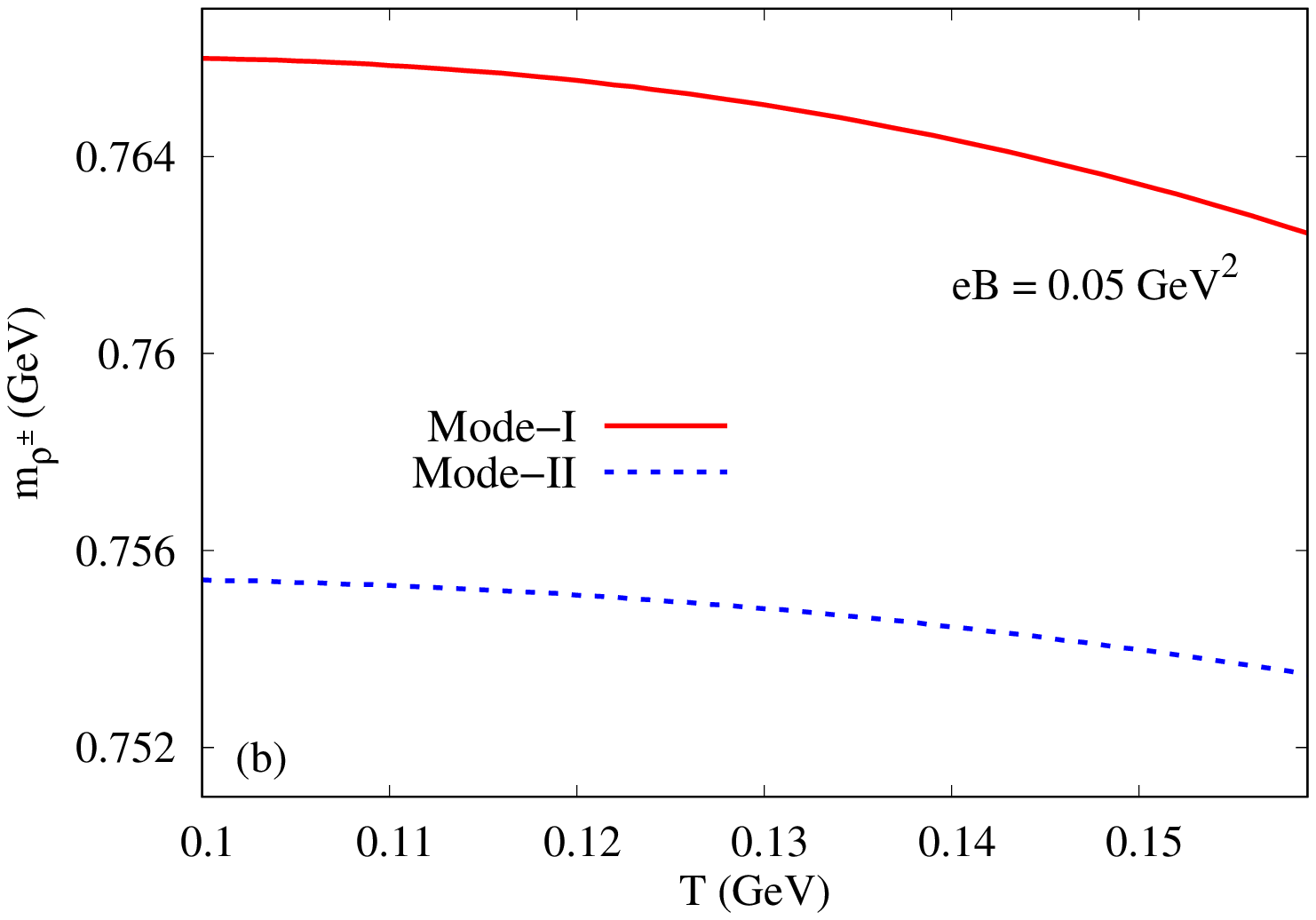}
	\caption{The effective masses of (a) $\rho^0$ and (b) $\rho^\pm$ vs $T$ for $eB=0.05$ GeV$^2$.}
	\label{Fig_rho_mass}	
\end{figure}
In Figs.~\ref{Fig_rho_mass} (a) and (b) we have shown the variation of effective masses of neutral and charged $\rho$ meson as a function of temperature for $ eB = 0.05 $ GeV$^2$. The masses have been calculated from the poles of the exact $\rho$-meson propagator in a thermo-magnetic medium~\cite{Ghosh:2017rjo}. It is well known that in thermo-magnetic medium there are three distinct propagating modes of vector meson. However at vanishing transverse momentum of the vector meson, two modes are identical and hence we are left with two distinct modes denoted by Mode-I and Mode-II respectively~\cite{Ghosh:2017rjo}. From both the figures it is clear that the change in effective masses of both neutral and charged meson with temperature is marginal. It should be noted that, the variation of mass with temperature/magnetic field calculated using the Lagrangian (Eq.~\eqref{eq.lagrangian}) in which $\rho$ and $\pi$ mesons are basic degrees of freedom is expected  to differ from that of other models such as Nambu--Jona-Lasinio (NJL) like models which are based on quark degrees of freedom and the mesons are generated by bosonization~\cite{Carlomagno:2022arc,Andreichikov:2016ayj,Ghosh:2020qvg,Luschevskaya:2016epp}.

\begin{figure}[h]
	\includegraphics[angle=-90, scale=0.44]{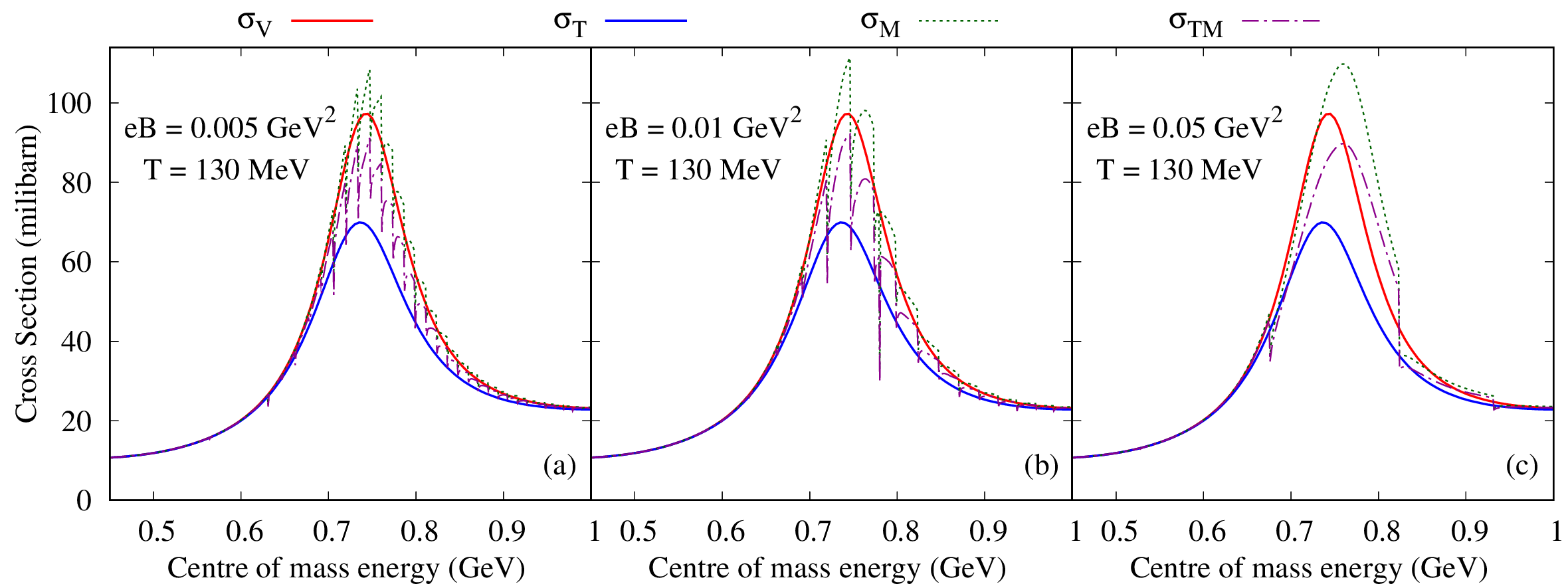}  
	\caption{(Colour Online) The variation of the isospin-averaged total $\pi\pi \rightarrow \pi\pi$ cross section at $T=130$ MeV as a function of the center-of-mass energy for different media at (a) $eB=0.005$ GeV$^2$, (b) $eB=0.01$ GeV$^2$ and (c) $eB=0.05$ GeV$^2$. The symbols $\sigma_\text{V}$, $\sigma_\text{M}$, $\sigma_\text{T}$ and $\sigma_\text{TM}$ denotes the $\pi\pi$ cross-section in vacuum, magnetic medium, thermal medium and thermo-magnetic medium respectively. } 
	\label{cross}
\end{figure}

We have plotted the $\pi\pi\rightarrow\pi\pi$ cross-section as a function of centre of mass energy for three different values of magnetic field- 0.005 GeV$^2$, 0.01 GeV$^2$ and 0.05 GeV$^2$ shown respectively in Fig.\ref{cross}(a), Fig.\ref{cross}(b) and Fig.\ref{cross}(c). In each of these plots temperature is taken to be 130 MeV. The numerical results of the cross-section has been computed in various media such as vacuum (V), thermal medium (T), magnetic medium (M), and thermo-magnetic medium (TM).

In Figs.~\ref{cross}(a) and (b) we observe the following trend for $\pi \pi$ cross-section: $\sigma_\text{M} > \sigma_\text{V} > \sigma_\text{TM} >\sigma_\text{T}$. We now attempt to explain this trend in the cross-section variation over different media. As the temperature is increased, the resonance spectral function experience in-medium broadening, owing to an increase in the imaginary part of the self-energy. This increase in the imaginary part of the self-energy in the thermal medium causes a decrease in the magnitude of thermal cross-section $\sigma_\tT$ compared to the vacuum cross-section $\sigma_\tV$. Physically, $\sigma_\tV>\sigma_\tT$ can be explained using the fact that in a thermal medium, $\rho$ meson suffer additional decay and scattering thus lowering the $\pi\pi$ cross-section in thermal medium. In all the figures we observe the presence of spikes for finite values of the background magnetic field. The spikes are purely due to the presence of magnetic field. Particularly, it is due to the combined effects of threshold singularity appearing in the imaginary part of $\rho^0$ self energy for each Landau level and the Laguerre polynomials (which produces oscillations) appearing in the imaginary part of $\rho^\pm$ self energy. Physically, it corresponds to highly unstable $\rho^0$ meson at threshold values of $q_0$ decaying to $\pi^{+}$ and $\pi^{-}$ in a purely magnetic medium. This explains the cross-section in magnetic medium being higher in magnitude compared to its vacuum counterpart ($\sigma_{\tM} > \sigma_{\tV}$). It is observed from the figure that the magnetic field causes nearly 10\% increase in magnitude of cross-section compared to the vacuum cross-section and the thermal bath causes approximately 25-30\% decrease in thermal medium dependent cross-section compared to the vacuum cross-section. This combined effect of temperature and magnetic-field causes the thermo-magnetic cross-section to be lower in magnitude than its vacuum counterpart as shown in Figs.~\ref{cross}(a) and (b). However, in Fig.~\ref{cross}(c) we observe the following trend - $\sigma_\text{M} > \sigma_\text{V} > \sigma_\text{TM} >\sigma_\text{T}$ for almost all values of centre of mass energy except for the region $0.75  \lesssim \sqrt{s} \lesssim 0.82$ GeV where $ \sigma_\text{TM} > \sigma_\text{V} $. This suggests that the effect of magnetic field dominates over thermal contribution near resonance energies at higher $eB$ values.
\begin{figure}[h]
	\includegraphics[angle=-90, scale=0.35]{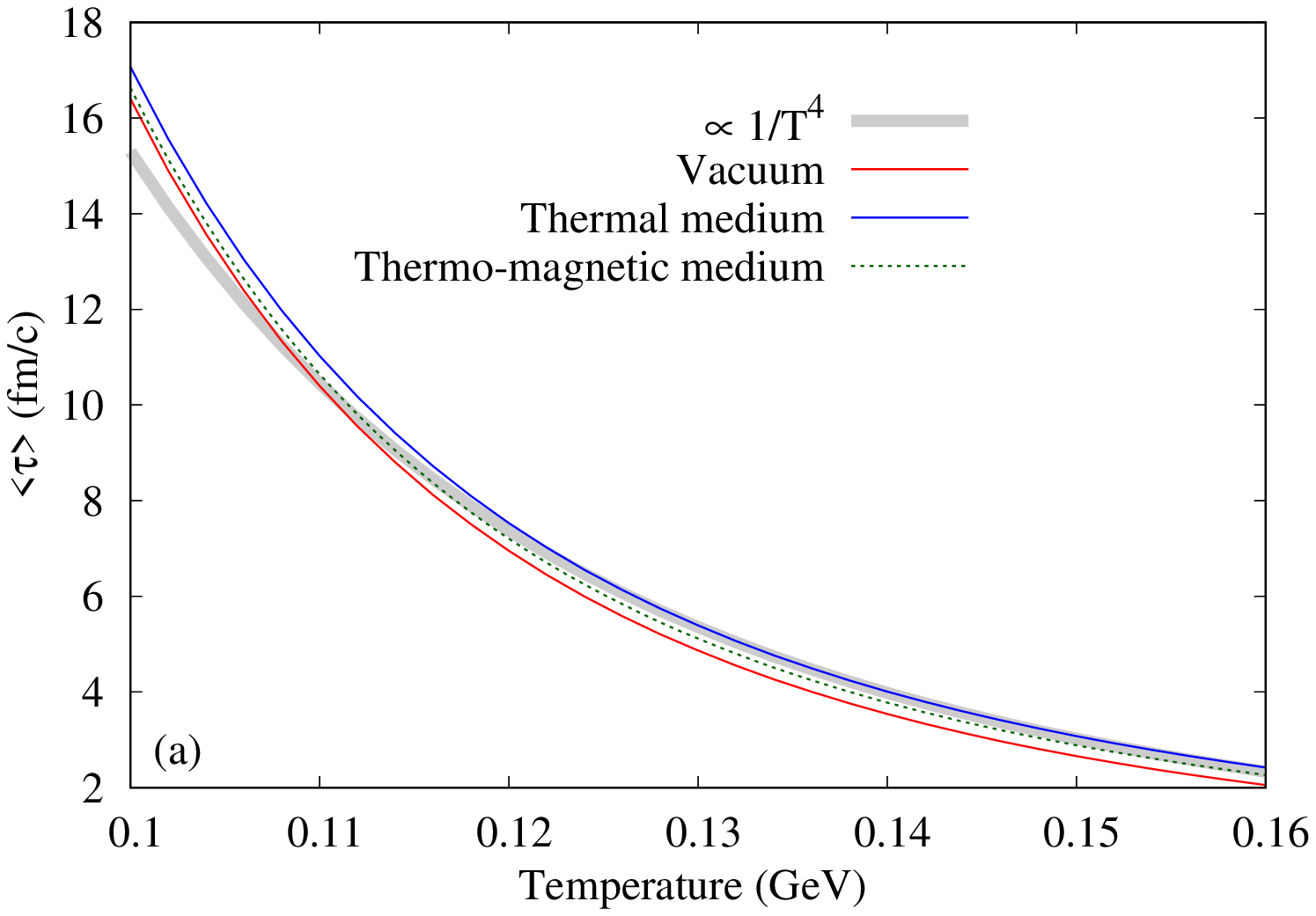}  
	\includegraphics[angle=-90, scale=0.35]{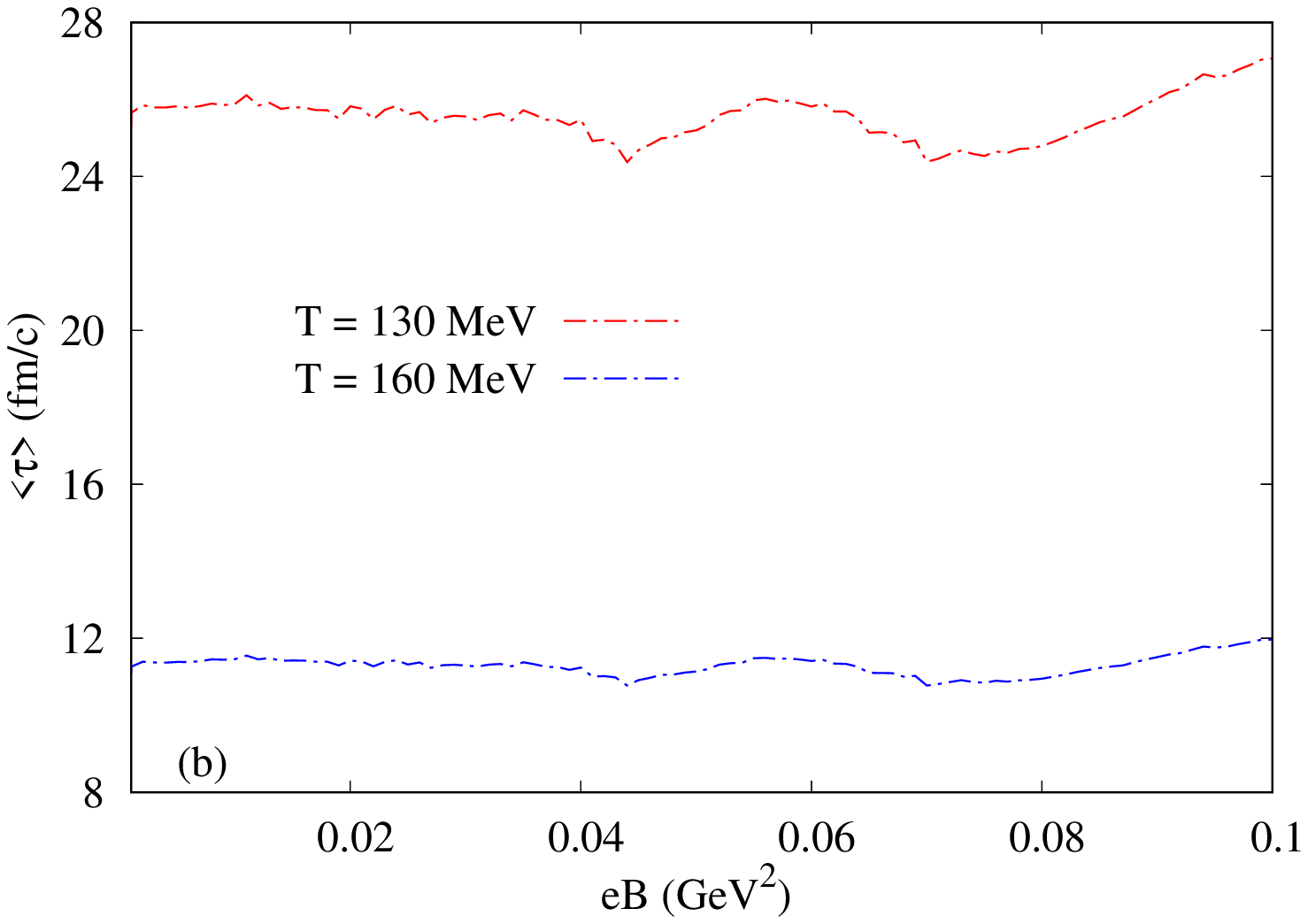} 
	\caption{(Colour Online) Variation of average relaxation time of pions as a function of (a) temperature at $eB=0.01$ GeV$^2$, and (b) magnetic field at different temperatures.} 
	\label{tau}
\end{figure}

Next, we have studied the variation of average relaxation time $\langle \tau \rangle $ of pions as a function of temperature at a magnetic field value of $eB=0.01$ GeV$^2$ in Fig.~\ref{tau} (a) and as a function of magnetic field at two different temperatures ($T=130$ MeV and $T=160$ MeV) in Fig.~\ref{tau}(b) respectively. In Fig.~\ref{cross}, it is seen that both the thermal and thermo-magnetic cross-section is suppressed with respect to the vacuum cross-section hence, the average relaxation time which is inversely related to the cross-section is enhanced in magnitude compared to its vacuum counterpart. As the relaxation time is a key dynamical input to the transport coefficients we will try to extract the leading behaviour of $\langle \tau \rangle $ as a function of temperature. For this we have fitted the  variation of average relaxation time with $\frac{a_0}{T^4}$ (here $a_0$ is a constant) for qualitative understanding of the results of electrical conductivity and shear viscosity. Thus the leading order behaviour of the average relaxation will be approximated to $ \langle \tau \rangle \sim \frac{1}{T^4}$. In Fig.~\ref{tau}(b), we observe a mild oscillatory variation of average relaxation time with respect to the magnetic field. As discussed earlier, this mild oscillatory behaviour is due to the combined effects of the threshold singularity appearing in the imaginary part of $\rho^0$ self-energy and Laguerre polynomial appearing in the $\rho^\pm$ self-energy.
\begin{figure}[h]
	\includegraphics[angle=-90, scale=0.35]{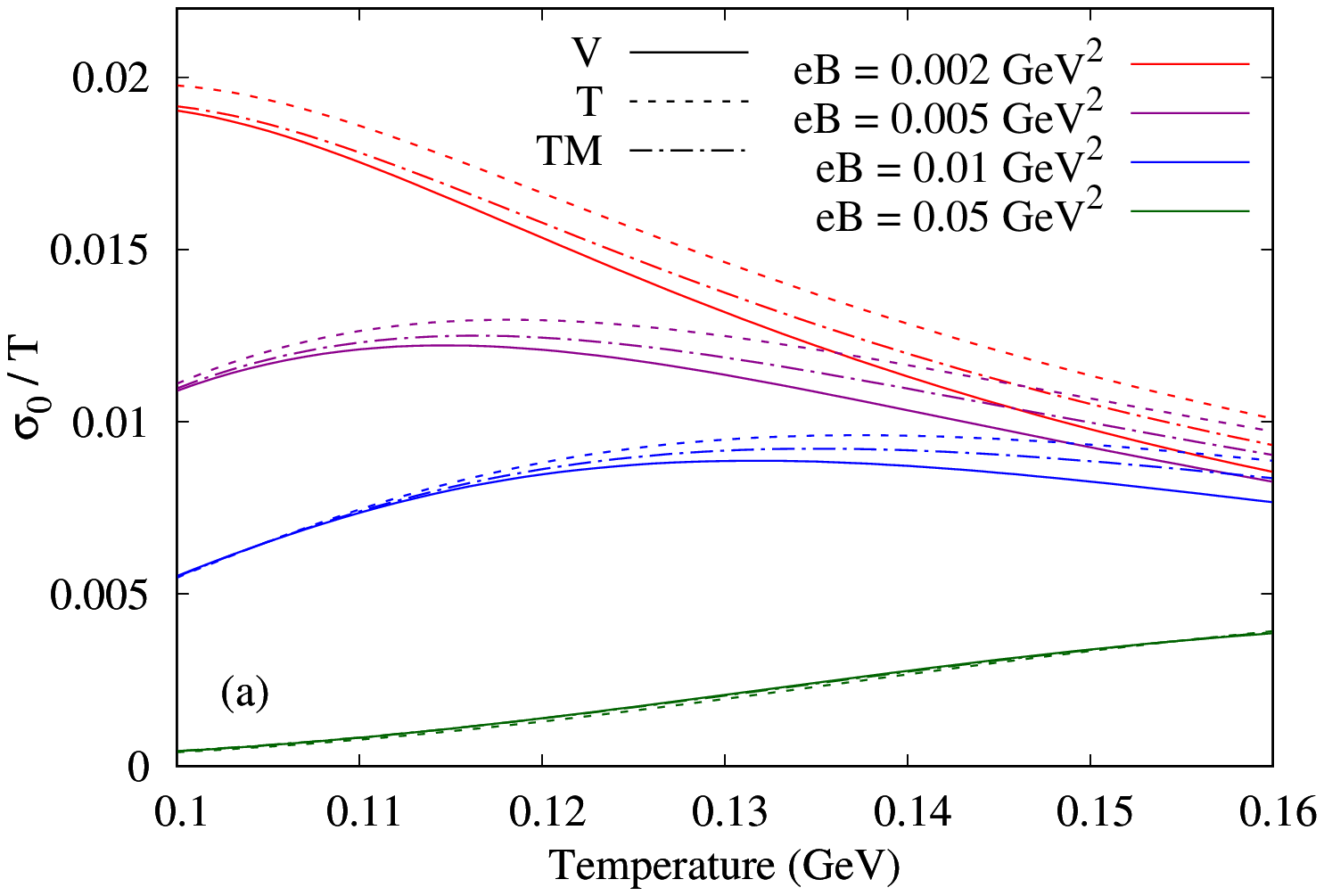}
	\includegraphics[angle=-90, scale=0.35]{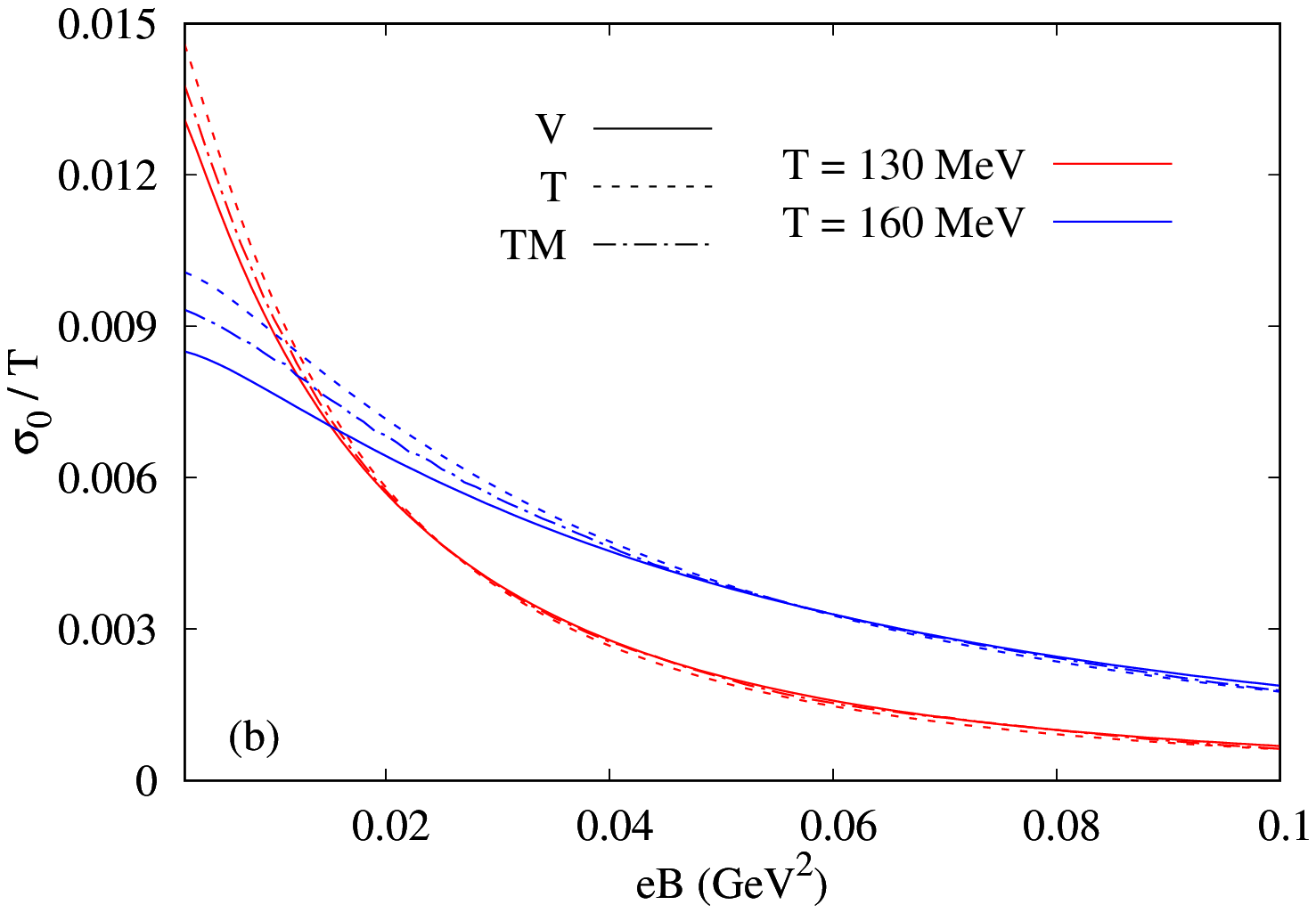} 
	\caption{(Colour Online) Variation of $\sigma_0/T$ as a function of (a) temperature for different values of the magnetic field strength, and (b) magnetic field for different values of the temperature. Solid, dashed and dash-dotted lines of different colours respectively represent the consideration of vacuum, thermal and thermo-magnetic cross-sections while calculating the transport coefficients.}
	\label{elec_cond}
\end{figure}

In Fig.~\ref{elec_cond}(a) we have shown the variation of $\frac{\sigma_0}{T}$ with temperature for different values of magnetic field strength, and in Fig.~\ref{elec_cond}(b) we have shown the variation of  $\frac{\sigma_0}{T}$ as a function of magnetic field for different values of temperature. It must be noted that the transport coefficients calculated using thermal medium dependent cross-section experiences magnetic field influence only through cyclotron frequency $\omega_c$, whereas in a thermo-magnetic medium magnetic field influence comes from $\omega_c$ and $eB$ dependent $\langle \tau \rangle $ calculated using medium dependent cross-section. The variation of $\frac{\sigma_0}{T}$ with temperature can be explained with $\frac{\sigma_0}{T} \sim \frac{\tau T}{1 + (\omega_c \tau)^2}$. We will make use of $\langle \tau \rangle \sim \frac{1}{T^4}$ to explain the temperature variation of electrical conductivity. We now explain the variation of electrical and Hall conductivities with temperature and magnetic field. For lower $eB$ values $\omega_c\tau \ll 1$, thus $\frac{\sigma_0}{T}\sim \tau T \sim \frac{1}{T^3}$ whereas for higher $eB$ values $\omega_c \tau \gg 1$, hence $\frac{\sigma_0}{T} \sim \frac{T}{\tau} \sim T^5$. This $\frac{1}{T^3}$ and $T^5$ variation respectively at lower and higher $eB$ values can be seen in Fig.~\ref{elec_cond}(a). The magnetic field dependence of $\frac{\sigma_0}{T}$ can be explained with $\frac{\sigma_0}{T} \sim \frac{\tau T}{1 + (\omega_c \tau)^2}$. The two magnetic field  dependent terms are $\omega_c$ and $\tau$. As $eB$ value increases, $\omega_c$ increases which results in the decrease of $\frac{\sigma_0}{T}$ for higher $eB$ values as seen in Fig.~\ref{elec_cond}(b).  For lower $eB$ values $\frac{\sigma_0}{T}\sim \tau$, thus we observe the trend $(\frac{\sigma_0}{T})_{\tT} > (\frac{\sigma_0}{T})_{\tTM} > (\frac{\sigma_0}{T})_{\tV} $ which is in agreement to the trend in average relaxation time $\langle \tau \rangle_{\tT} > \langle \tau \rangle_{\tTM} > \langle \tau \rangle_{\tV}$. However, this trend in $\frac{\sigma_0}{T}$ variation with magnetic field is not respected for higher $eB$ values as $\frac{\sigma_0}{T} \sim \frac{1}{\tau}$ for higher $eB$ values.
\begin{figure}[h]
	\includegraphics[angle=-90, scale=0.35]{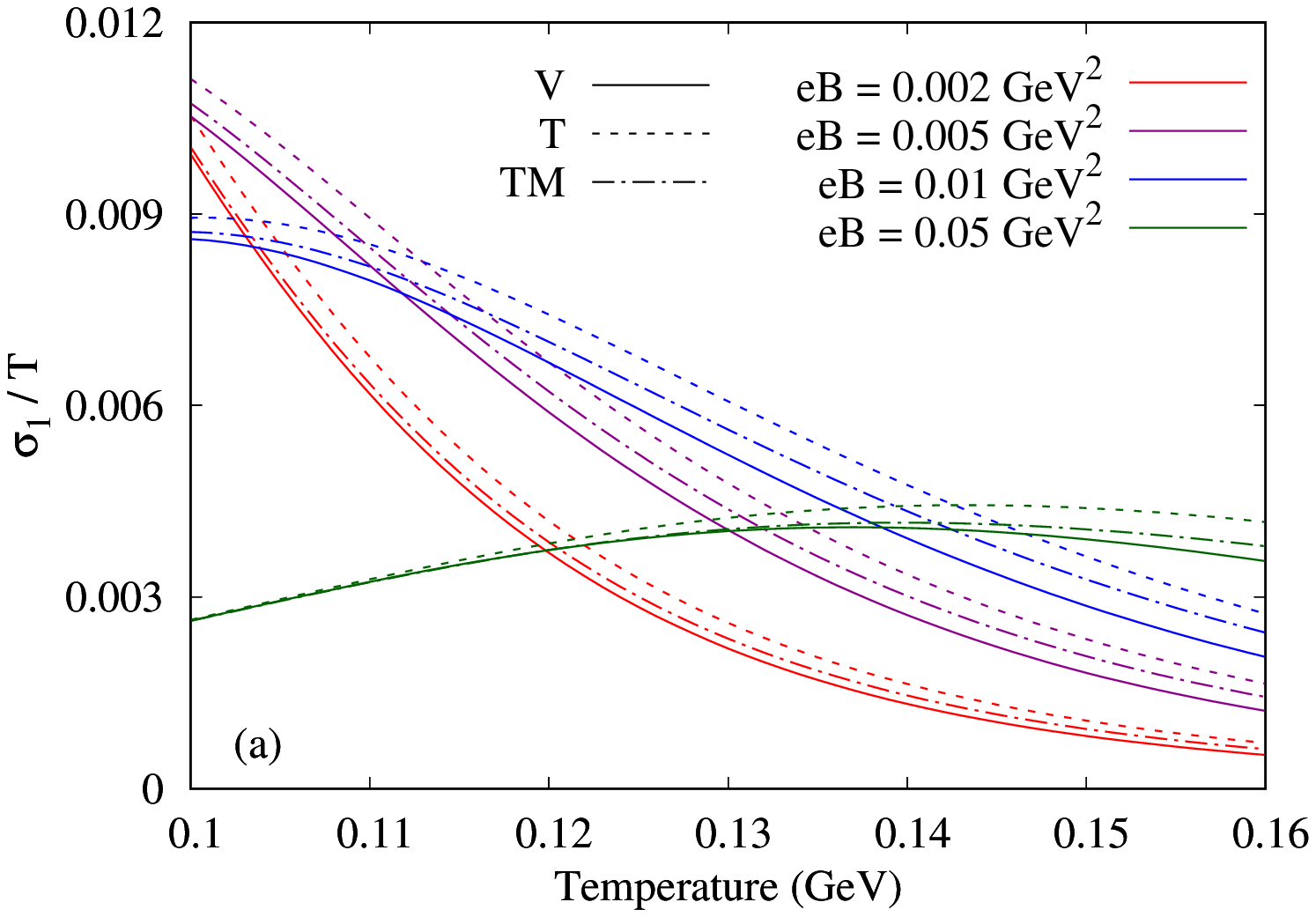}  
	\includegraphics[angle=-90, scale=0.35]{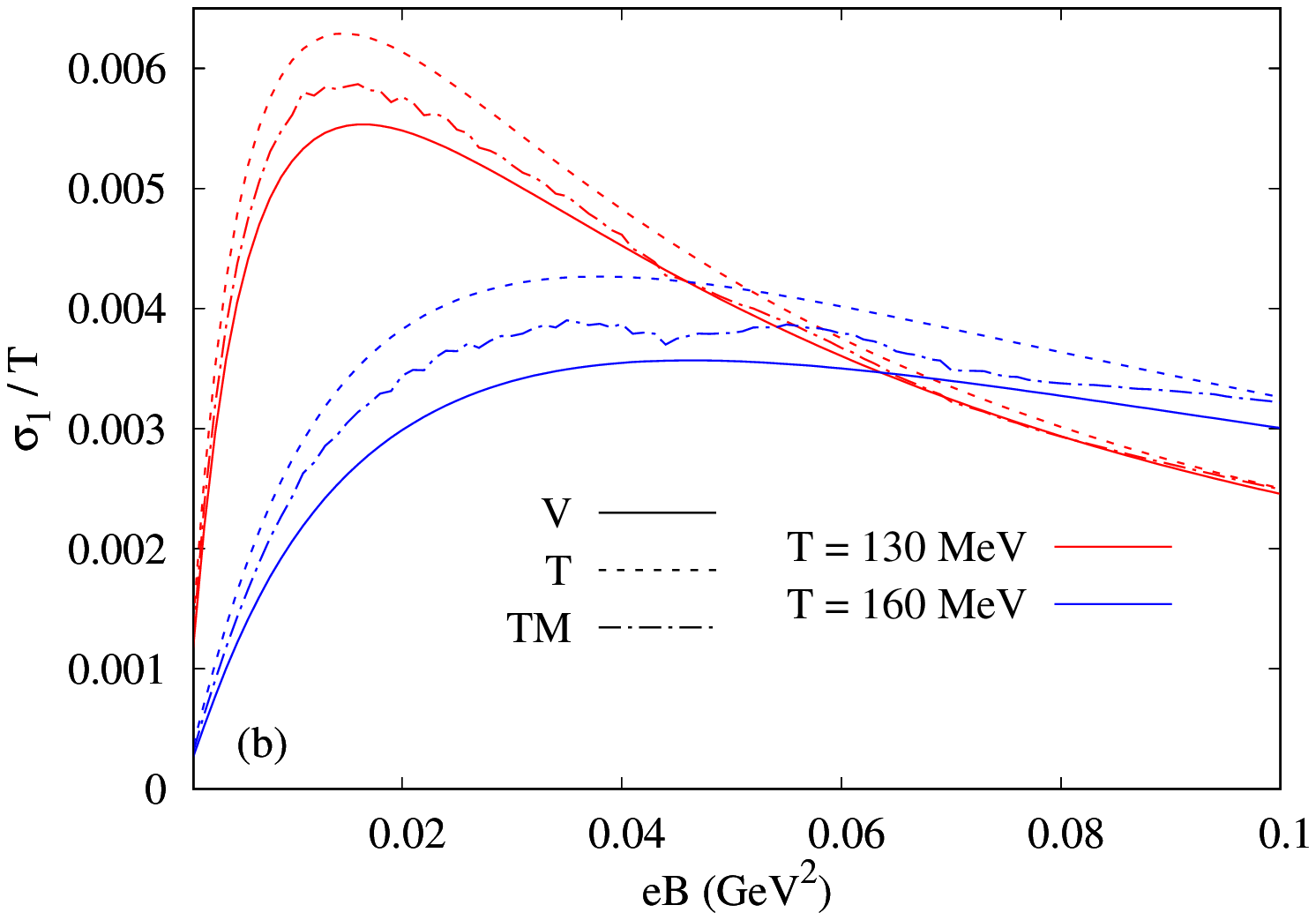} 
	\caption{(Colour Online) Variation of $\sigma_1/T$ as a function of (a) temperature for different values of the magnetic field strength, and (b) magnetic field for different values of the temperature. Solid, dashed and dash-dotted lines of different colours respectively represent the consideration of vacuum, thermal and thermo-magnetic cross-sections while calculating the transport coefficients.}
	\label{hall_cond}
\end{figure}

In Fig.~\ref{hall_cond}(a) we have shown the plot for the variation of temperature scaled Hall conductivity $\frac{\sigma_1}{T}$ as a function of temperature for different values of magnetic field strength and in Fig.~\ref{hall_cond}(b), we have shown the variation of $\frac{\sigma_1}{T}$ as a function of magnetic field for different values of temperature. The temperature variation of $\frac{\sigma_1}{T}$ can be understood with $\frac{\sigma_1}{T} \sim \frac{\tau^2 \omega_c T}{1 + (\omega_c \tau)^2}$. For lower $eB$ values $\omega_c \tau \ll 1$, which results in $\frac{\sigma_1}{T}\sim \tau^2\omega_c T \sim \frac{\omega_c}{T^7}$ whereas for higher $eB$ values $\omega_c \tau \gg 1$, resulting in $\frac{\sigma_1}{T}\sim \frac{T}{\omega_c}$. This variation can be observed in Fig.~\ref{hall_cond}(a). The variation of $\frac{\sigma_1}{T}$ with magnetic field as seen in Fig.~\ref{hall_cond}(b) can be understood using $\frac{\sigma_1}{T} \sim \frac{\tau^2 \omega_c}{1 + (\omega_c \tau)^2}$. The cyclotron frequency, $\omega_c$ renders a Breit-Wigner like function and $\tau^2$ causes mild oscillations in $\frac{\sigma_1}{T}$ for a thermo-magnetic medium due to the $eB$ dependent $\pi\pi$ scattering cross-section. As the medium effect information is contained in the relaxation time $\tau$ and as $\frac{\sigma_1}{T}\sim \tau^2$, we observe in Figs.~\ref{hall_cond}(a) and (b) that for any value of temperature and magnetic field, $\frac{\sigma_1}{T}$ evaluated for the medium (thermal and thermo-magnetic both) is higher than its vacuum counterpart.
\begin{figure}[h]
	\includegraphics[angle=-90, scale=0.35]{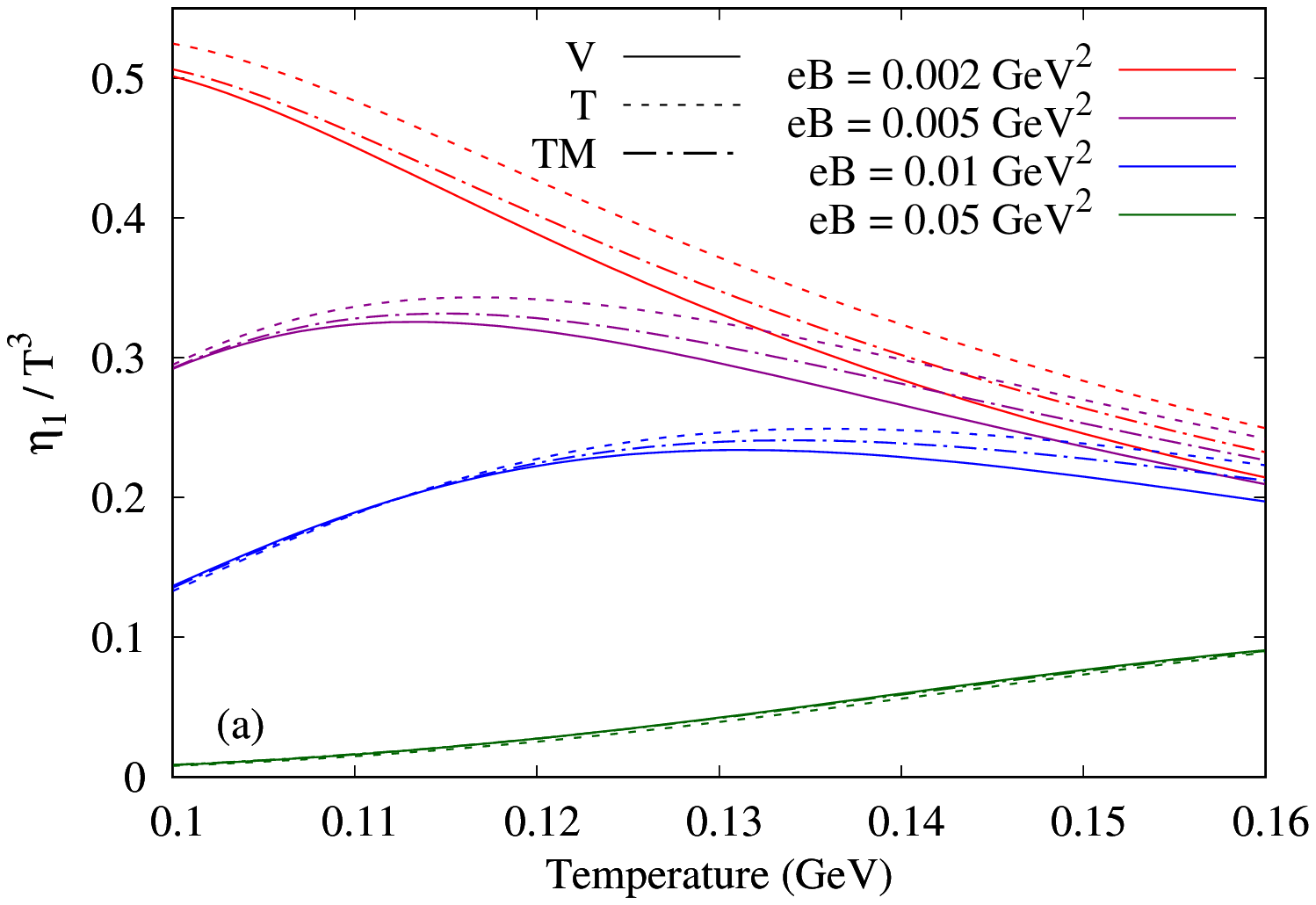}
	\includegraphics[angle=-90, scale=0.35]{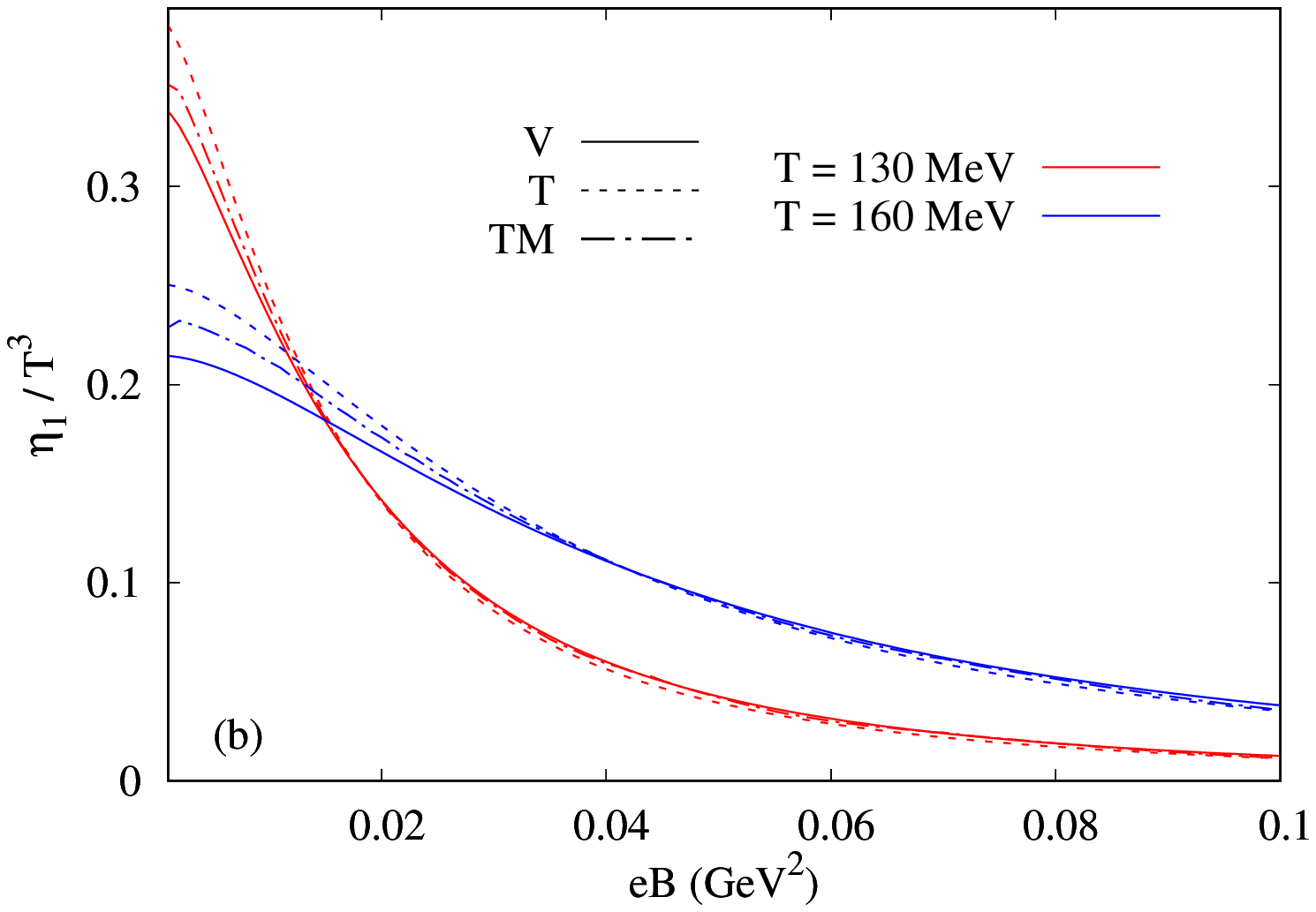}
	\caption{(Colour Online) Variation of $\eta_1/T$ as a function of (a) temperature for different values of the magnetic field strength, and (b) magnetic field for different values of the temperature. Solid, dashed and dash-dotted lines of different colours respectively represent the consideration of vacuum, thermal and thermo-magnetic cross-sections while calculating the transport coefficients.}
	\label{eta1}
\end{figure}
\begin{figure}[h]
	\includegraphics[angle=-90, scale=0.35]{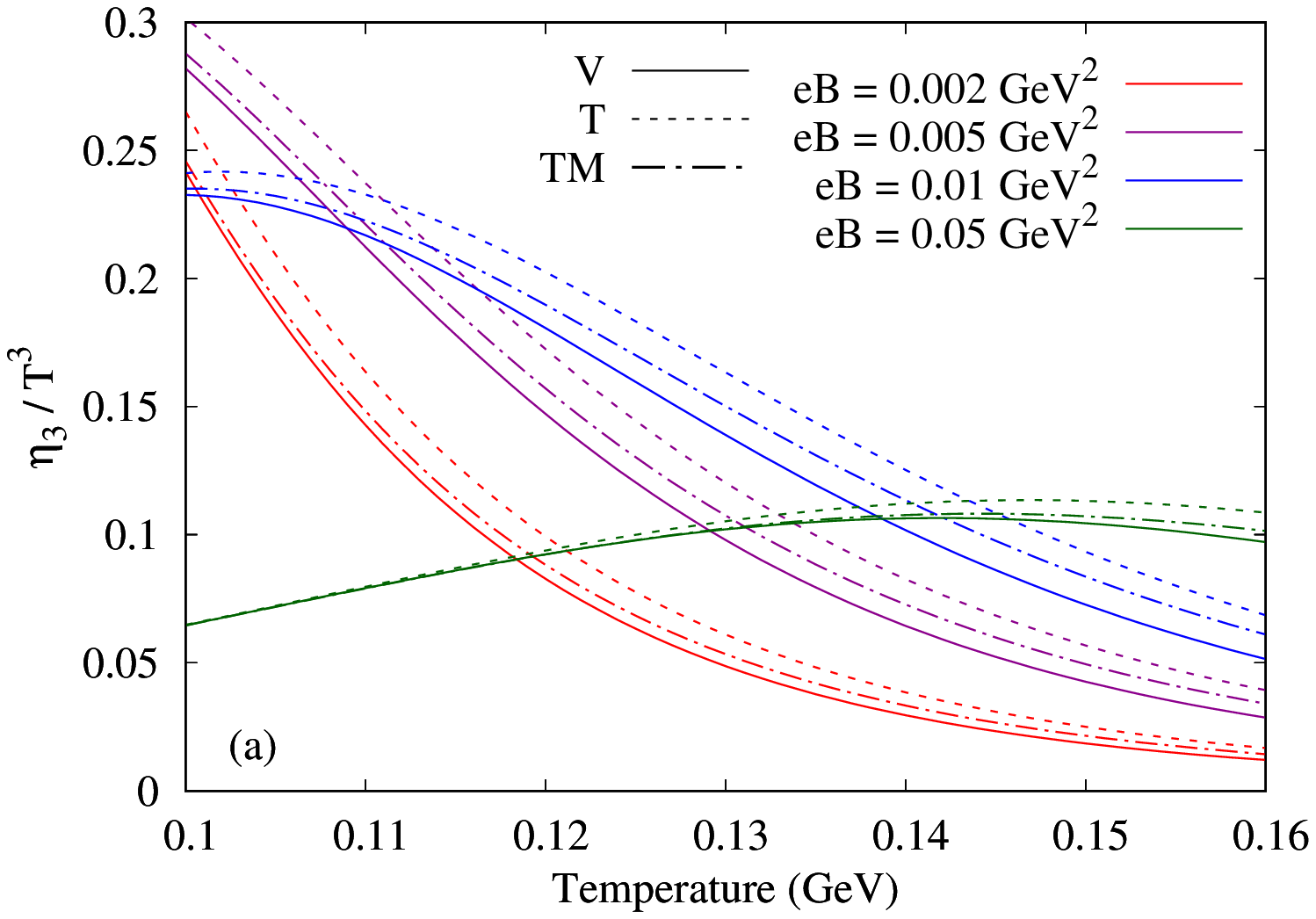}
	\includegraphics[angle=-90, scale=0.35]{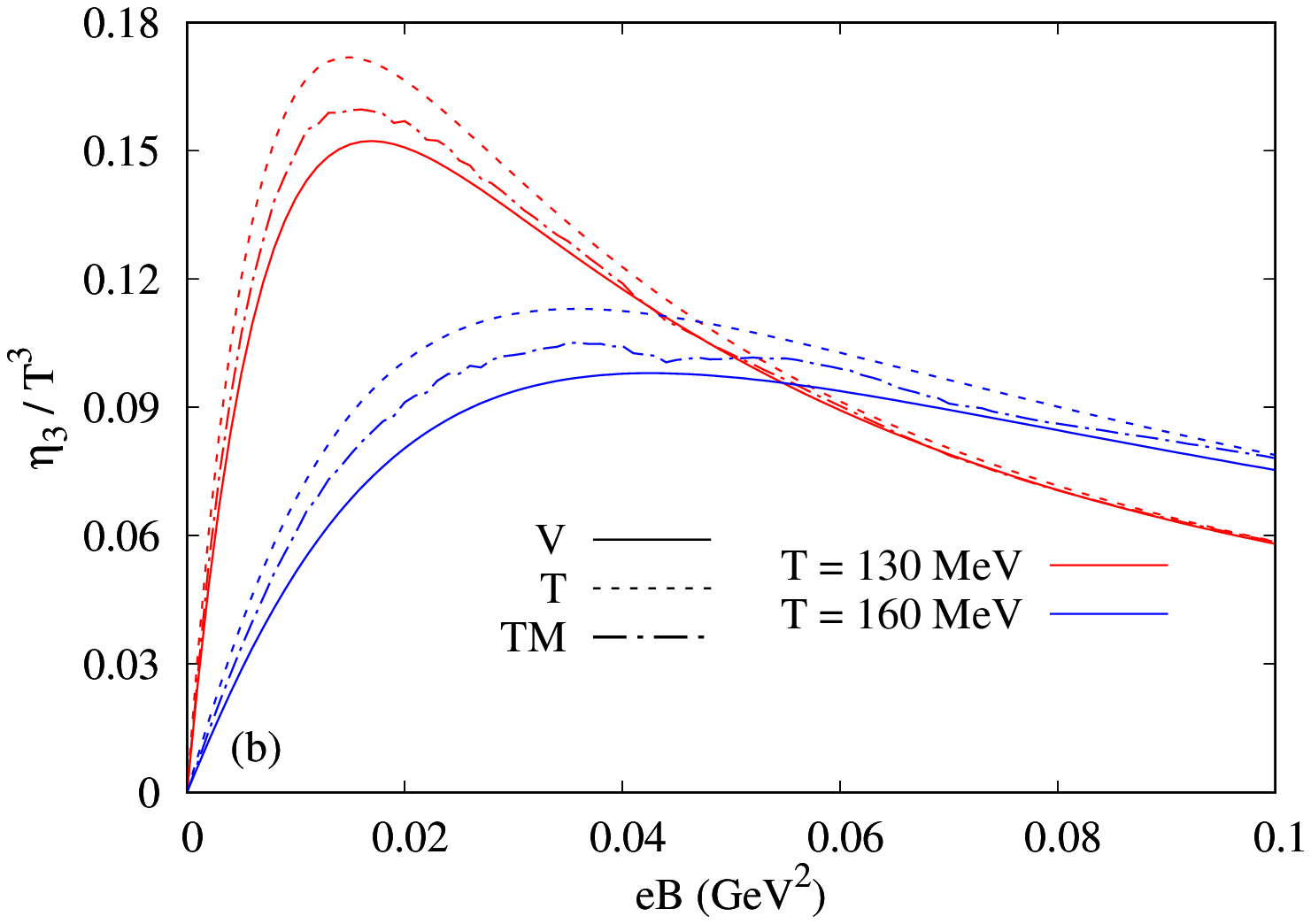}
	\caption{(Colour Online) Variation of $\eta_3/T$ as a function of (a) temperature for different values of the magnetic field strength, and (b) magnetic field for different values of the temperature. Solid, dashed and dash-dotted lines of different colours respectively represent the consideration of vacuum, thermal and thermo-magnetic cross-sections while calculating the transport coefficients.}
	\label{eta3}
\end{figure}

The variation of temperature scaled normal shear viscosity $\frac{\eta_1}{T^3}$ with temperature for different values of magnetic field is shown in Fig.~\ref{eta1}(a) and the variation of $\frac{\eta_1}{T^3}$  with magnetic field for different values of temperature is shown in Fig.~\ref{eta1}(b). Using $\frac{\eta_1}{T^3} \sim \frac{\tau T}{1 +  (2 \omega_c \tau)^2}$, we can explain the temperature dependence of $\frac{\eta_1}{T^3}$. For $\omega_c\tau \ll 1$ corresponding to lower $eB$ values, $\frac{\eta_1}{T^3}\sim \tau T \sim \frac{1}{T^3}$ whereas for $\omega_c\tau \gg 1$ corresponding to higher $eB$ values, $\frac{\eta_1}{T^3}\sim \frac{T}{\tau}\sim T^5$. At intermediate magnetic field values the variation of $\frac{\eta_1}{T^3}$ is non-monotonic which is due to the interplay of both temperature and $eB$. The variation of $\frac{\eta_1}{T^3}$ with magnetic field can be understood using $\frac{\eta_1}{T^3}\sim \frac{\tau}{1 + (2\omega_c \tau)^2}$. As $\ensembleaverage{\tau}$ is approximately constant (mild oscillations) with changing $eB$ as seen in Fig.~\ref{tau}(b), the $\frac{\eta_1}{T^3}$ variation with magnetic field can be explained using the values of $\omega_c$. Thus, with the increase in $\omega_c$ corresponding to higher $eB$ values, $\frac{\eta_1}{T^3}$ decreases monotonically. As observed earlier $\langle \tau\rangle_{\tT}>\langle \tau\rangle_ {\tTM}>\langle \tau\rangle_{\tV}$ and for lower $eB$ values $\frac{\eta_{1}}{T^3}\sim \tau$, we observe the trend $(\frac{\eta_1}{T^3})_{\tT} > (\frac{\eta_1}{T^3})_{\tTM} > (\frac{\eta_1}{T^3})_{\tV}$ for lower $eB$ values in Fig.~\ref{eta1}(a) and (b). The variation of $\frac{\eta_2}{T^3}$ with $eB$ and temperature can be explained similar to that of $\frac{\eta_1}{T^3}$ as they differ only by a factor in the denominator.
\begin{figure}[h]
	\includegraphics[angle=-90, scale=0.35]{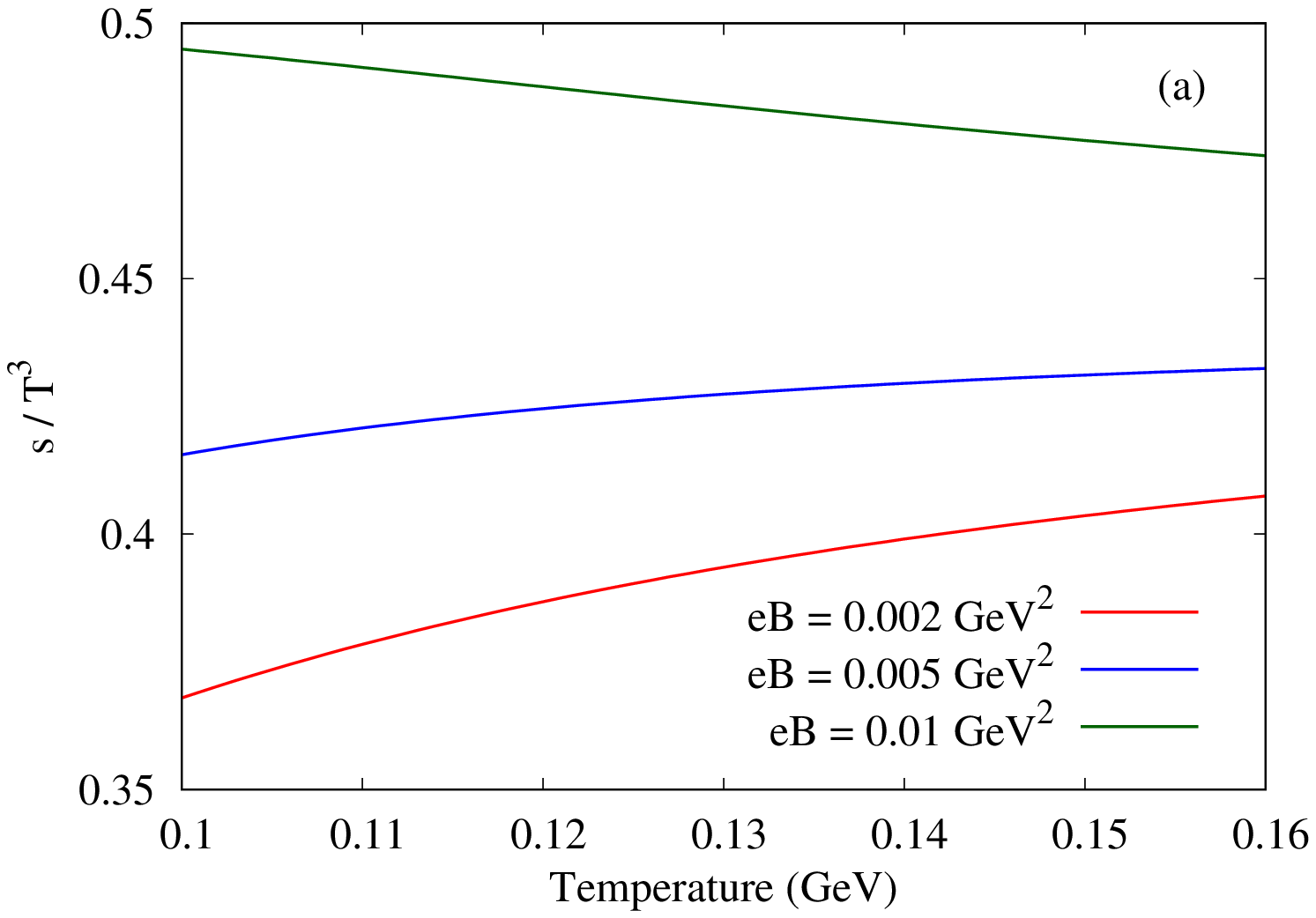}
	\includegraphics[angle=-90, scale=0.35]{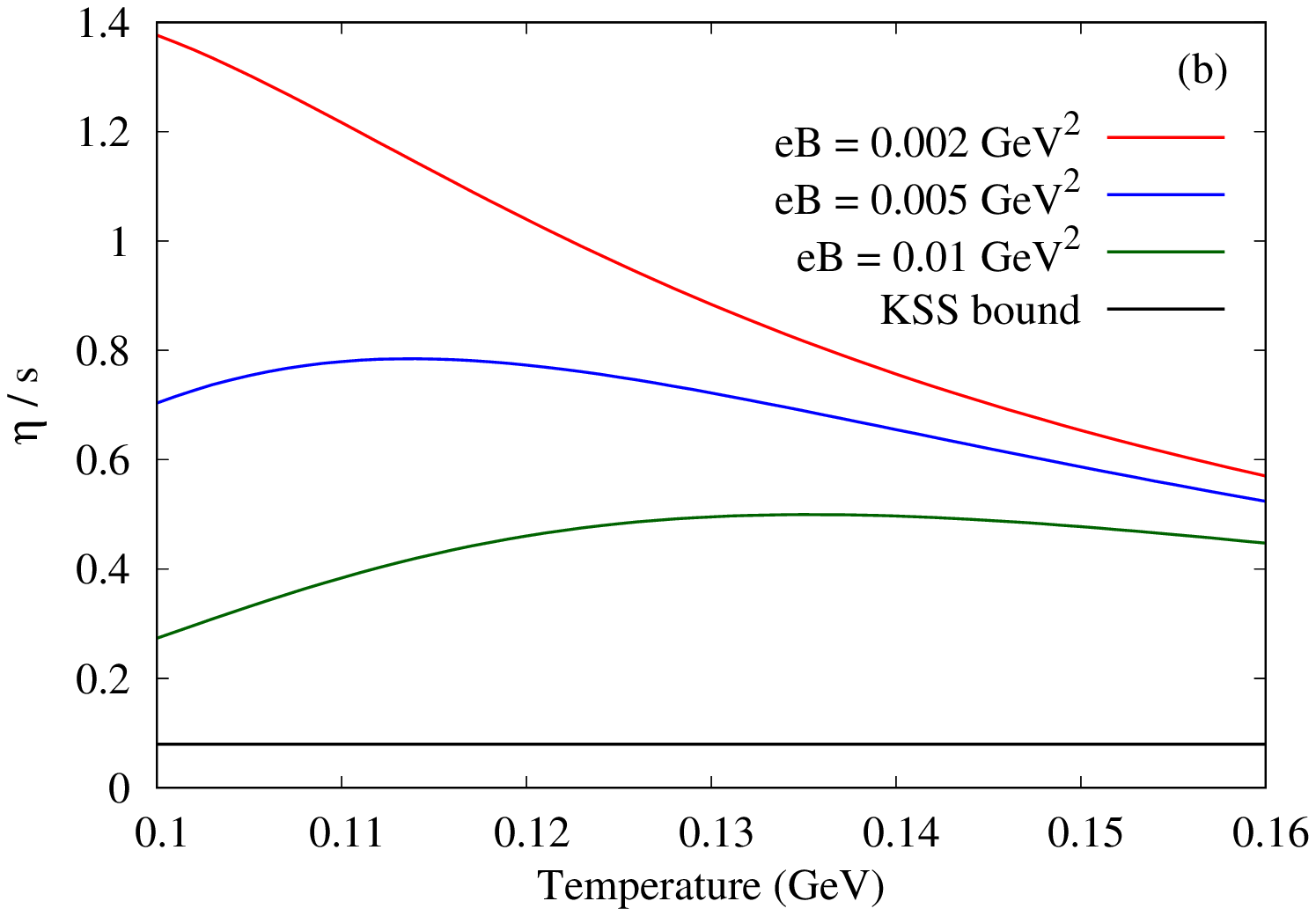} 
	\caption{(Colour Online) Variation of (a) entropy density ($s$), and (b) Specific shear viscosity ($\eta/s$) as a function of temperature for different values of the magnetic field strength.} \label{entropy_density}
\end{figure}

In Fig.~\ref{eta3}(a) we have shown the plot for the variation of temperature scaled Hall type shear viscosity  $\eta_3/T^3$ as a function of temperature for different values of magnetic field whereas in Fig.~\ref{eta3}(b) we have shown the variation of $\eta_3/T^3$ as a function of magnetic field for different values of temperature. $\eta_3$ is a purely Hall type quantity. It vanishes at zero magnetic field as can be seen from Fig.~\ref{eta3}(b). The variation of $\frac{\eta_3}{T^3}$ with temperature can be understood with $\frac{\eta_3}{T^3} \sim \frac{\tau^2 \omega_c T}{1 + (2\omega_c \tau)^2 }$. At low $eB$ values $\omega_c\tau \ll 1$, thus $\frac{\eta_3}{T^3} \sim \tau^2 \omega_c T \sim \frac{\omega_c }{T^7}$ whereas for higher $eB$ values $\omega_c\tau \gg 1$, thus $\frac{\eta_3}{ T^3} \sim \frac{T\tau^2}{\tau^2 \omega_c T}\sim \frac{T}{\omega_c }$. This causes $\frac{\eta_3}{T^3}$ to monotonically decrease with increasing temperature at lower $eB$ values and to monotonically increase with increasing temperature at higher $eB$ values. The variation of $\frac{\eta_3}{T^3}$ with $eB$ can be explained using $\frac{\eta_3}{T^3} \sim \frac{\tau^2 \omega_c}{1 + (2\omega_c\tau)^2}$. The quantity $\tau^2$ appearing in the numerator has an effect of causing mild oscillations in $\frac{\eta_3}{T^3}$ calculated for the thermomagnetic medium. A Breit-Wigner like structure seen in Fig.~\ref{eta3}(b) is due to the term $\frac{\omega_c}{1 + (2\omega_c\tau)^2}$ occurring in $\frac{\eta_3}{T^3}$. Both Breit-Wigner form and mild oscillations in thermo-magnetic medium can be observed in Fig.~\ref{eta3}(b). It can been observed from Figs.~\ref{eta3}(a) and (b) that $\frac{\eta_3}{T^3}$ calculated in thermal and thermo-magnetic medium has higher magnitude for all values of $eB$ compared to that of $\frac{\eta_3}{T^3}$ evaluated in the vacuum with finite $B$. This is in agreement with the fact that $\eta_3$ is a Hall type quantity and increasing magnetic field necessarily increases its value. The variation of $\frac{\eta_4}{T^3}$ as a function of temperature and magnetic field can be explained similar to that of $\frac{\eta_3}{T^3}$ since their expressions differ only by a factor in the denominator as can be noticed by comparing Eqs.~\eqref{eta_3} and \eqref{eta_4}.
\begin{figure}[h]
	\includegraphics[angle=-90, scale=0.35]{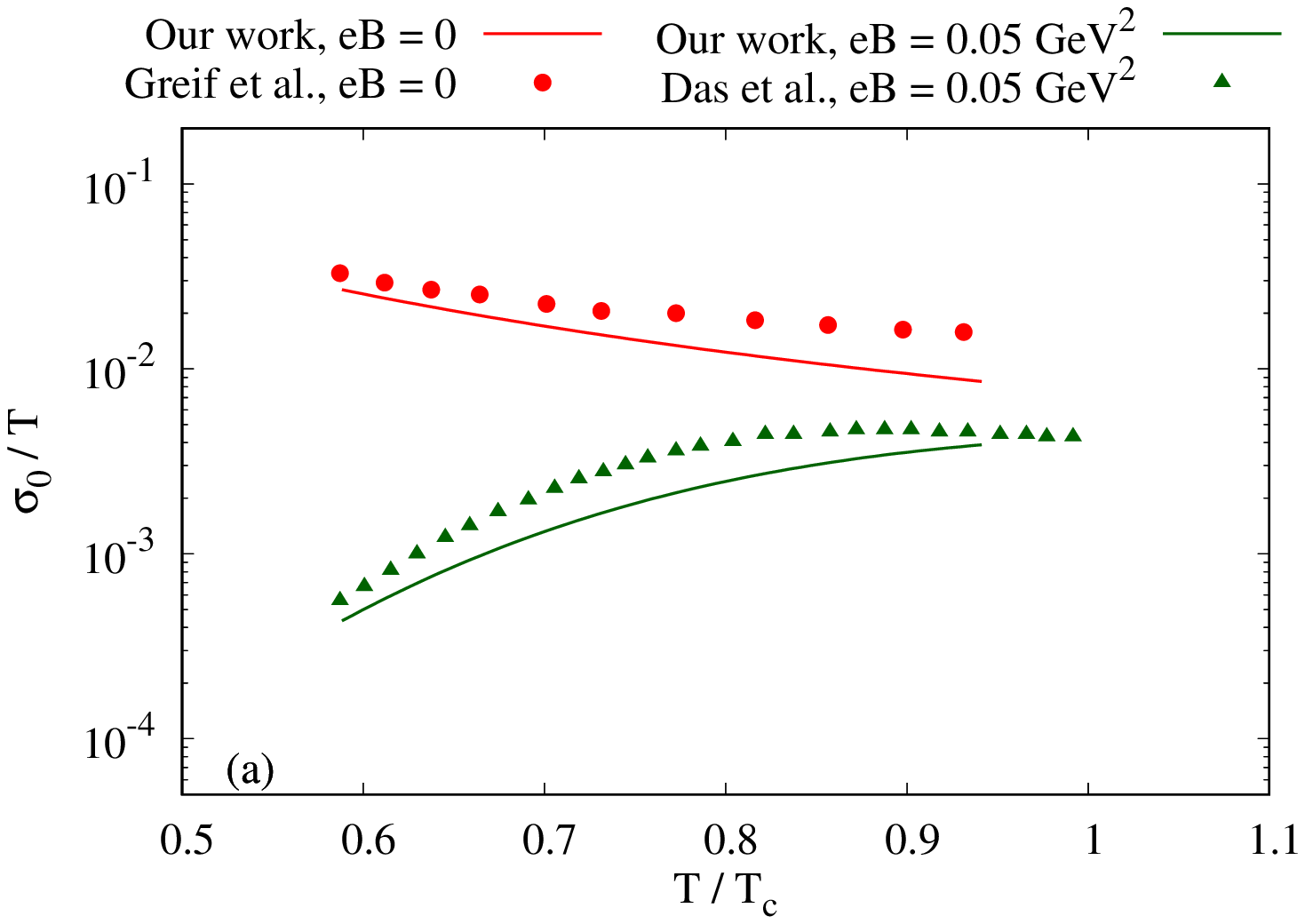}
	\includegraphics[angle=-90, scale=0.35]{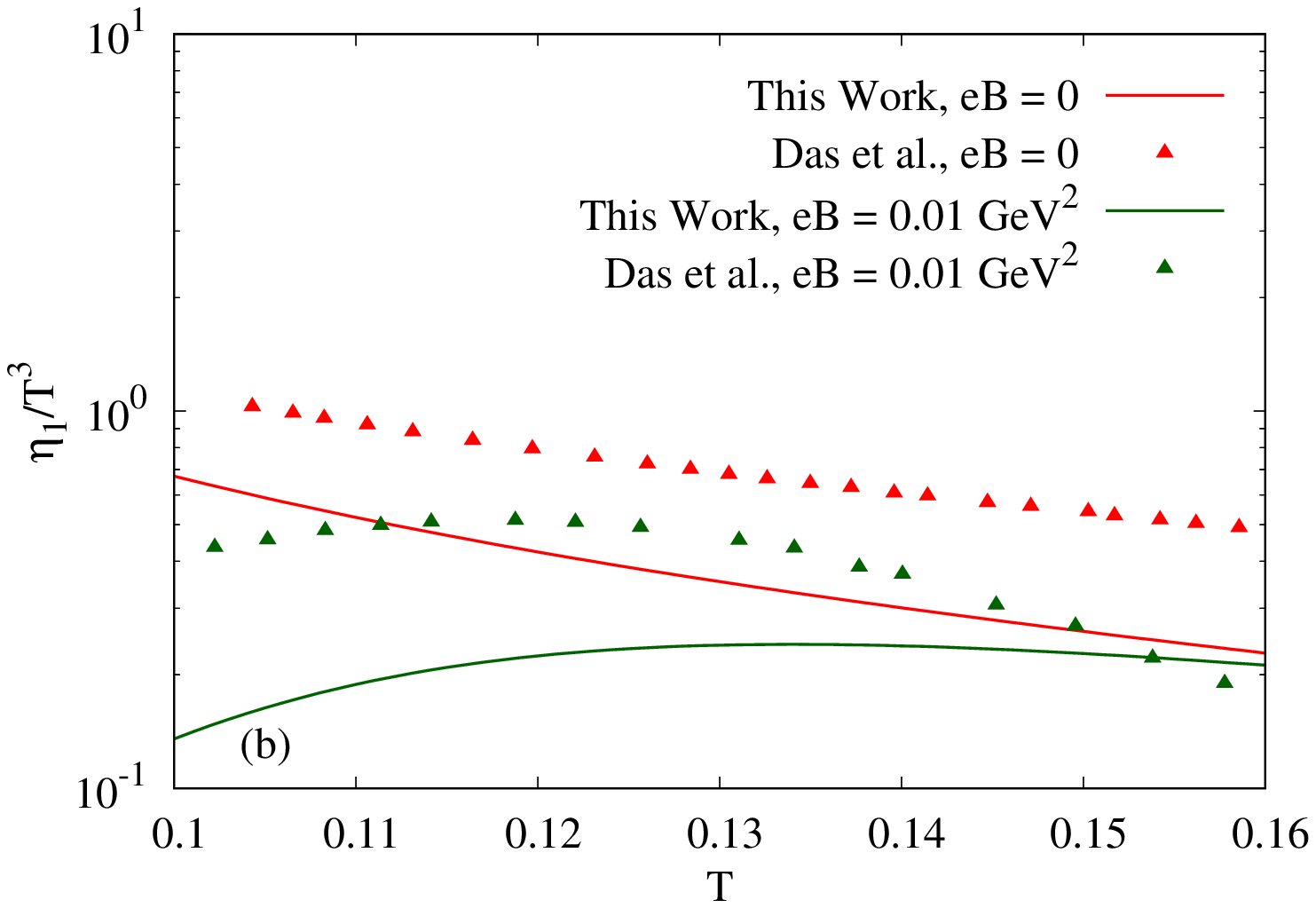}
	\caption{(Colour Online) Comparison of (a) $\sigma_0/T$ obtained in this work at zero magnetic field with Refs.~\cite{Amato:2013naa} and \cite{Greif:2016skc} and at non-zero magnetic field with Refs.~\cite{Feng:2017tsh} and \cite{Das:2019ppb}, and (b) $\eta_1/T^3$ obtained in this work with at zero and non-zero magnetic field with Ref.~\cite{Das:2019pqd} } 
	\label{comparison_plot}
\end{figure}

In Fig.~\ref{entropy_density}(a), we have shown the variation of scaled entropy density with temperature. The entropy density $s$ has been calculated using the thermodynamic potential as shown in Appendix~\ref{app.entropy}. It is seen from the figure that for lower values of magnetic field $s/T^3$ increases with increasing temperature whereas for higher values of magnetic field strength $s/T^3$ decreases with increasing temperature. This shows that the increasing magnetic field reduces the randomness in pionic system under consideration. In Fig.~\ref{entropy_density}(b), we have shown the variation of the specific shear viscosity ($\eta/s$) as a function of temperature for different values of magnetic field in a thermo-magnetic medium.  As the magnitude of entropy density increases with the increase in magnetic field for a given temperature, it causes an increase in the magnitude of $\eta/s$ compared to the magnitude of $\eta_1/T^3$. The obtained values of $\eta/s$ are well within the KSS bound.

In Fig.~\ref{comparison_plot}(a), we have made a comparison of electrical conductivity obtained in our work with the other available results in literature. For zero magnetic field our results show a good match with the works of Greif et al. in Ref.~\cite{Greif:2016skc}. Electrical conductivity results for $eB=0.05 GeV^2$  agrees well with the electrical conductivity calculated by Das et al. at the same $eB$ value for a hadron resonance gas in Ref.~\cite{Das:2019ppb}. Similarly in Fig.~\ref{comparison_plot}(b), we have shown a comparison of shear viscosity for both zero and non-zero magnetic field with that obtained by Das et al. in Ref.~\cite{Das:2019pqd}. The difference in magnitude of shear viscosity can be attributed to the increased number density in a hadron resonance gas (used by Das et al.) which causes an increase in the magnitude of shear viscosity in Ref.~\cite{Das:2019pqd}.
%~~~~~~~~~~~~~~~~~~~~~~~~~~~~~~~~~~~~~~~~~~~~~~~~~~~~~~~~~~~~~~~~~~~~~~~~~~~~
%
\section{Summary} \label{sec.sum}
In this work we have evaluated the electrical conductivity and shear viscosity of a pion gas in a thermo-magnetic medium using kinetic theory and relaxation time approximation. The dynamics of the pion gas in a thermo-magnetic medium is taken into consideration by evaluating the $\pi\pi \rightarrow \pi\pi$ scattering cross-section  using thermal field theory techniques in presence of background magnetic field. A medium dependent relaxation time obtained from the medium dependent cross-section has been used to calculate the electrical conductivity and shear viscosity of the pion gas. 

The magnetic field influences the electrical conductivity and shear viscosity through cyclotron frequency and the magnetic field dependent relaxation time. It is observed that the relaxation time shows a mild oscillatory behaviour with magnetic field which is reflected in the electrical conductivity and shear viscosity evaluated in a thermo-magnetic medium. However, it must be noted that the magnetic field influence coming from the cyclotron frequency is higher in magnitude than the magnetic field influence from the relaxation time.

Electrical conductivity and shear viscosity show an increase in magnitude with respect to temperature for a thermo-magnetic medium compared to its vacuum counterpart. Electrical conductivity $\sigma_0/T$ and shear viscosity $\eta_1/T^3$ for a particular value of temperature show increased magnitude for lower magnetic field values and the magnitude gradually decreases for higher magnetic field values when thermo-magnetic effects are considered. However, the Hall type  electrical conductivity $\sigma_1/T$ and Hall type shear viscosity $\eta_3/T^3$ for a particular value of temperature is found to increase for all values of the magnetic field. 

The electrical conductivity and shear viscosity obtained in this work shows a good agreement with the works available in the literature. Also, we have observed that our estimated value of electrical conductivity in the range of 0.5-2 MeV causes a maximum magnetic field of strength $10^{-6}\text{-}10^{-5}$ GeV$^2$ to survive for a time $t\sim 10$ fm which is sufficient to affect the dynamics and hence the evolution of hadronic matter produced in the later stage of HICs.

It is to be noted that we have not included the interesting phenomena of $ \rho^+-\pi^+ $ mixing~\cite{Carlomagno:2022arc} as observed in NJL type models. However such kind of mixing cannot be achieved in the simplistic model that we have used here. Also the neutral pion remains unaffected by the magnetic field in this case. Moreover, for $ eB=0.05 $ GeV$ ^2 $ the Landau magnetic length is of the order of the radius of $\rho$ and hence an appropriate form factor is necessary to suppress the manifestation of quark structure of the mesons  while calculating the cross-section. We have checked by using a few representative form factors including that of~\cite{Carlomagno:2022arc}, however no appreciable change in the results is observed.

\section*{Acknowledgments}
S.G. is funded by the Department of Higher Education, Government of West Bengal, India. 
%~~~~~~~~~~~~~~~~~~~~~~~~~~~~~~~~~~~~~~~~~~~~~~~~~~~~~~~~~~~~~~~~~~~~~~~~~~~~~
%
\appendix
\section{Tensors Appearing in the Imaginary Parts of Self-Energies}\label{app.tensors}
In this appendix, we provide the explicit expressions of the tensors $U_{1,nl}^\munu$, $U_{2,nl}^\munu$, $L_{1,nl}^\munu$, and $L_{2,nl}^\munu$ appearing in Eq.~\eqref{impi0} and $U_{1,l}^\munu$, $U_{2,l}^\munu$, $L_{1,l}^\munu$, and $L_{2,l}^\munu$ appearing in Eq.~\eqref{impipm}. These are given by
\begin{eqnarray}
	U_{1,nl}^\munu( q^0,k_z) &=& \SB{ 1 + f_0(\omega_k^n) + f_0(\omega_k^l) } N^{\mu\nu}_{nl}(k_0=\omega_k^n),\\
	U_{2,nl}^\munu( q^0,k_z) &=&  \SB{ -1-f_0(\omega_k^n) - f_0(\omega_k^l)} N^{\mu\nu}_{nl}(k_0=-\omega_k^n), \\
	L_{1,nl}^\munu( q^0,k_z) &=&  \SB{ f_0(\omega_k^n) - f_0(\omega_k^l)} N^{\mu\nu}_{nl}(k_0=-\omega_k^n), \\	
	L_{2,nl}^\munu( q^0,k_z) &=&  \SB{ -f_0(\omega_k^n) + f_0(\omega_k^l)} N^{\mu\nu}_{nl}(k_0=\omega_k^n),
	\\
%\end{eqnarray}
%\begin{eqnarray}
	U_{1,l}^\munu\big(q^0,|\bm{k}|,\cos\theta\big) &=& 2 (-1)^{l} e^{-\alpha_k} L_l(2\alpha_k) \SB{1 + f_0(\omega_k^l) + f_0(\wkv)} N^{\mu\nu}(q,k^0 = \wkv,\bm{k}), \\
	U_{2,l}^\munu\big(q^0,|\bm{k}|,\cos\theta\big) &=& 2 (-1)^{l} e^{-\alpha_k} L_l(2\alpha_k) \SB{ -1 - f_0(\omega_k^l) - f_0(\wkv) } N^{\mu\nu}(q,k^0 =-\wkv,\bm{k}), \\
	L_{1,l}^\munu\big(q^0,|\bm{k}|,\cos\theta\big) &=& 2 (-1)^{l} e^{-\alpha_k} L_l(2\alpha_k) \SB{ f_0(\omega_k^l) - f_0(\wkv)} N^{\mu\nu}(q,k^0 = -\wkv,\bm{k}), \\
	L2_{1,l}^\munu\big(q^0,|\bm{k}|,\cos\theta\big) &=& 2 (-1)^{l} e^{-\alpha_k} L_l(2\alpha_k) \SB{ -f_0(\omega_k^l) + f_0(\wkv)} N^{\mu\nu}(q,k^0 = \wkv,\bm{k}),
\end{eqnarray}
where,
\begin{eqnarray}
	\mathcal{N}^{\mu\nu}_{nl}(q^0,k^0,k_z) &=& g_{\rho\pi\pi}^2(-1)^{n+l}
	\left(\frac{ eB}{2\pi}\right) \Big[   \left\{q_\parallel^4k_\parallel^\mu k_\parallel^\nu +
	(q_\parallel.k_\parallel)^2q_\parallel^\mu q_\parallel^\nu-q_\parallel^2(q_\parallel.k_\parallel)(q_\parallel^\mu
	k_\parallel^\nu+q_\parallel^\nu k_\parallel^\mu)\right\}\delta^n_l \nn \\
	&& \hspace{2cm}  -~\frac{eB}{4}q_\parallel^4g_\perp^{\mu\nu}\left\{ (2n+1)\delta^n_l-n\delta^{n-1}_l-(n+1)\delta^{n+1}_l\frac{}{} \right\}\Big], \label{N_munu_nl}
\end{eqnarray}
$\omega_k^n = \sqrt{k_z^2 + m_n^2}$ and $N^\munu$ is defined below Eq.~\eqref{Pi.11.pm}.
%
%~~~~~~~~~~~~~~~~~~~~~~~~~~~~~~~~~~~~~~~~~~~~~~~~~~~~~~~~~~~~~~~~~~~~~~~~~~~~~~~~~~~~~~~~~~~~~~~~~~~~~~~~~~~~~~~~~~~~~~~~~~~~~~~~~~~~~
\section{Calculation of Entropy Density}\label{app.entropy}
In order to derive the expression for entropy density for a system of pion gas, we make use of the thermodynamic potential following~\cite{Strickland:2012vu,Andersen:2014xxa,Endrodi:2013cs,Chaudhuri:2020lga,Chaudhuri:2022oru,Atta:2022wxs,Avancini:2020xqe}. The normalized thermodynamic potential $\Omega$ in the absence of magnetic field is given as
\begin{eqnarray}
	\Omega^{\rm norm}(T) = g T\int\!\!\! \frac{d^3 k}{(2\pi)^3} \ln(1-e^{-\wkv/T})
\end{eqnarray}
where $\wkv = \sqrt{\bm{k}^2 + m_\pi^2}$ and $g=3$ is the pion degeneracy. Here the normalization is done in the usual manner by subtracting the divergent vacuum contribution. The other relevant thermodynamic quantities can be calculated using the normalized thermodynamic potential as
\begin{eqnarray}
	P&=&-\Omega^{\rm norm}(T), \\
	\epsilon &=& -T^2 \frac{\partial }{\partial T} \left(\frac{\Omega^{\rm norm}}{T}\right)
\end{eqnarray}
where $P$ is the pressure and $\epsilon$ is the energy density. The entropy density is then obtained from $s=\FB{\dfrac{\epsilon +P}{T}}$.

The presence of background magnetic field can be incorporated by modifying the momentum integral for charged pions in the following manner~\cite{Andersen:2014xxa}
\begin{eqnarray}
	\int\!\!\! \frac{d^3 k}{(2\pi)^3} f(\omega_k) \longrightarrow \sum_{l=0}^{\infty} \frac{|eB|}{2\pi} \int_{-\infty}^{\infty} \frac{d k_z}{2\pi} f(\omega^l_k).
\end{eqnarray}
Here $\omega_k^l(k_z)= \sqrt{k_z^2 + (2l+1)eB + m_\pi^2} $ is the Landau quantized energy eigen value of the charged pions. Thus the normalized thermodynamic potential for a pion gas in presence of external magnetic field is then given by
\begin{eqnarray}
	\Omega^{\rm norm}_{B}(T,eB) =  T\int\!\!\! \frac{d^3 k}{(2\pi)^3} \ln(1-e^{-\wkv/T})  + 2 T\sum_{l=0}^{\infty} \frac{|eB|}{2\pi} \int_{-\infty}^{\infty} \frac{d k_z}{2\pi} \ln(1-e^{-\omega_k^l/T})
\end{eqnarray}
where, the first term on the rhs is the contribution from the neutral pions which are not affected by the magnetic field.
The entropy density in presence of magnetic field is given by~\cite{Strickland:2012vu,Andersen:2014xxa}
\begin{equation}
	 s_{B}=\frac{\epsilon_{B} + P_{\parallel}}{T}
\end{equation}
where, $P_{\parallel}= -\Omega^{\rm norm}_{B}$ is the longitudinal pressure and 
\begin{eqnarray}
	\epsilon_{B}= \frac{1}{3}\epsilon + 2 \sum_{l=0}^{\infty} \frac{|eB|}{2\pi} \int_{-\infty}^{\infty} \frac{d k_z}{2\pi} \frac{\omega_k^l} {e^{\omega_k^l/T}-1}.
\end{eqnarray}

\bibliography{pallavi}

\end{document}